\documentclass[a4paper,fleqn,usenatbib]{mnras}

\usepackage{newtxtext,newtxmath}

\usepackage[T1]{fontenc}
\usepackage{ae,aecompl}

\usepackage{graphicx}	
\usepackage{amsmath}	
\usepackage{amssymb}	
\usepackage{graphicx}	
\usepackage{amsmath}	
\usepackage{amssymb}	
\usepackage[shortlabels]{enumitem}      
\usepackage{svg}    
\usepackage{booktabs}   
\usepackage{algorithm}  
\usepackage[noend]{algpseudocode}  
\makeatletter
\def\BState{\State\hskip-\ALG@thistlm}
\makeatother

\definecolor{orange}{rgb}{0.8,0.35,0}
\definecolor{purple}{rgb}{0.3,0.15,0.55}
\newcommand{\orange}[1]{#1}
\newcommand{\purple}[1]{#1}
\newcommand{\pquad}{\hspace{0.05cm}}

\graphicspath{{Figures}}
\DeclareGraphicsExtensions{.png,.pdf,.eps}

\defcitealias{Semelin2017}{21SSD}
\defcitealias{Karras2017}{PGGAN}

\title[21cm tomography with PGGANs]{A unified framework for 21cm tomography sample generation and parameter inference with Progressively Growing GANs}

\author[F. List \& G. F. Lewis]{
Florian List\thanks{E-mail: flis0155@uni.sydney.edu.au} and
Geraint F. Lewis
\\
Sydney Institute for Astronomy, School of Physics, A28, The University of Sydney, NSW 2006, Australia
}

\date{Accepted XXX. Received YYY; in original form ZZZ}

\pubyear{2019}

\begin{document}
\label{firstpage}
\pagerange{\pageref{firstpage}--\pageref{lastpage}}
\maketitle

\begin{abstract}
    Creating a database of 21cm brightness temperature signals from the Epoch of Reionisation (EoR) for an array of reionisation histories is a complex and computationally expensive task, given the range of astrophysical processes involved and the possibly high-dimensional parameter space that is to be probed. 
    We utilise a specific type of neural network, a Progressively Growing Generative Adversarial Network (PGGAN), to produce realistic tomography images of the 21cm brightness temperature during the EoR, covering a continuous three-dimensional parameter space that models varying X-ray emissivity, Lyman band emissivity, and ratio between hard and soft X-rays. The GPU-trained network generates new samples at a resolution of $\sim 3'$ in a second (on a laptop CPU), and the resulting global 21cm signal, power spectrum, and pixel distribution function agree well with those of the training data, taken from the 21SSD catalogue \citep{Semelin2017}. Finally, we showcase how a trained PGGAN can be leveraged for the converse task of inferring parameters from 21cm tomography samples via Approximate Bayesian Computation.
    
\end{abstract}


\section{Introduction}
The 21cm spectral line holds a great potential for probing the Dark Ages, the formation of the first generations of stars, and the Epoch of Reionisation 
\citep[EoR; see][for reviews of 21cm cosmology]{Furlanetto2006, Morales2010, Pritchard2012}.
It is produced by a spin-flip transition of H$_\text{I}$ gas in the intergalactic medium (IGM) due to the absorption of Cosmic Microwave Background (CMB) photons. This signal from the early Universe, which is received on Earth highly redshifted in the radio frequencies, is a dormant treasure trove that is expected to reveal a plethora of information on inflationary and dark sector physics \citep{Furlanetto2019a}, the matter power density after recombination \citep{Loeb2004}, and the reionisation history of the Universe \citep{Furlanetto2019}.
\par Since the signal is concealed by foreground sources that can be $\sim$ 5 magnitudes brighter \citep{Pober2013, Dillon2014}, its detection is extremely challenging, and it is only recently that the first tentative absorption profile from the Dark Ages has been measured in the global (sky-averaged) signal by EDGES \citep[\emph{Experiment to Detect the Global EoR Signature},][]{Bowman2018}. Currently operating interferometers such as LOFAR \citep[\emph{Low Frequency Array},][]{VanHaarlem2013}, GMRT \citep[\emph{Giant Metrewave Radio Telescope,}][]{Paciga2013}, MWA \citep[\emph{Murchison Widefield Array},][]{Tingay2013}, and PAPER \citep[\emph{Precision Array for Probing the Epoch of Reionization},][]{Parsons2010} are not sensitive enough to map the spatial fluctuations of the 21cm signal and aim instead at characterising the signal by means of statistics such as the power spectrum \citep{Mellema2015}, for which several \orange{upper limits} at different redshifts have been derived \citep[e.g.][]{Beardsley2016, Patil2017, Barry2019, Eastwood2019, Gehlot2019, Kolopanis2019, Li2019}. However, future radio interferometer experiments such as the SKA \citep[\emph{Square Kilometer Array},][]{Mellema2013} and HERA \citep[\emph{Hydrogen Epoch of Reionization Array},][]{DeBoer2017} will usher in an era of 21cm imaging. 

\par With these powerful telescopes coming online within the next few years, accurate and fast emulators are needed that generate templates for a wide range of astrophysical parameters and reionisation models. The EDGES signal shows a trough with an amplitude more than twice as large as can be accommodated by the most extreme predictions \citep{Cohen2017}, which, if confirmed, could hint at an excess radio background (e.g. \citealt{Feng2018, Fialkov2019}) from a so far unknown population of high-redshift radio sources such as black holes \citep{Ewall-Wice2018} or supernovae (\citealt{Jana2019}, but see \citealt{Sharma2018} for constraints on the cooling time when resorting to astrophysical explanations), or at exotic physics such as scattering between dark matter and baryons \citep{Barkana2018, Hirano2018, Slatyer2018}. Whether or not this signal will be substantiated by other instruments, predicting signals for all physically plausible models and conducting parameter studies is vital in order to be prepared for upcoming observations.
However, generating mock samples is a challenging task: the 21cm signal depends both on cosmology and on astrophysical processes, which are not well understood to date. Moreover, running computationally expensive large-scale cosmological simulations with radiative transfer \orange{requires} both big volumes as well as high resolutions in order to resolve the ionising sources, \orange{which} often makes the exploration of a large parameter space infeasible. Approximate methods for the computation of the 21cm signal that do not invoke 3D radiative simulations have been developed, e.g. \textsc{21cmFast} \citep{Mesinger2011} and \textsc{SimFast21} \citep{Santos2010}, whereas methods such as \textsc{Bears} \citep{Thomas2009} post-process dark-matter-only simulations to generate 21cm maps.

\par Machine learning techniques lend themselves to vastly speeding up the creation of mock samples as well as carrying out fast and accurate parameter inference. In the following, we will briefly review the related literature: \citet{Kern2017} present an emulator for the 21cm power spectrum based on Gaussian processes and use it to forecast the constraining power of HERA. A neural-network-based power spectrum emulator is introduced in \citet{Schmit2018}, which can be harnessed for accelerating the evaluation of the likelihood function during Bayesian parameter inference; \orange{this present
paper shares the spirit of their approach
 in that the output of a generative model is used for parameter inference.} \citet{Chardin2019} predict the reionisation times in three-dimensional cubes using convolutional neural networks (CNNs). \citet{Cohen2019} generate smooth global 21cm signals with a neural network that is trained to predict the strength of the principal components depending on seven astrophysical parameters. A complementary approach to the work herein, which is based upon generative adversarial networks (GANs) as well, has been presented by \citet{Zamudio-Fernandez2019}. In that paper, a GAN is trained on 3D cubes from the IllustrisTNG simulation \citep{Pillepich2018} and learns to generate new realisations of the H$_\text{I}$ distribution. In terms of statistics, the quality of GAN-generated samples exceeds that of Halo Occupation Distributions. In contrast to modelling the \emph{sources} of 21cm emission, our method addresses the prediction of the signal as received on Earth.
\par While the aforementioned works consider the task of artificial data creation, the reverse problem of machine-learning-aided parameter inference has also been studied in the recent years. Pioneering the use of neural networks for the analysis of the 21cm signal, \citet{Shimabukuro2017} train a neural network on 21cm power spectra to infer astrophysical parameters. A comprehensive comparison of different machine learning methods for parameter estimation from the 21cm power spectrum is carried out in \citet{Doussot2019}. To date, the majority of machine learning contributions to 21cm cosmology has focused on the power spectrum, which is arguably the simplest and most powerful statistics when it comes to constraining EoR models. However, unlike the CMB, the 21cm signal is highly non-Gaussian (e.g. \citealt{Majumdar2018}), and valuable additional information is encoded in the full tomographic signal. In this spirit, \citet{Gillet2019} consider the recovery of astrophysical parameters by means of a neural network that is trained directly on 21cm images. \citet{LaPlante2019} determine the EoR duration from 2D images in the plane of the sky using a CNN within an error of $5$ per cent, assuming a well-constrained midpoint of the EoR. A CNN is also employed by \citet{Hassan2019}, who infer cosmological and astrophysical parameters from 21cm mock maps with SKA-like noise. A neural network classifier that discerns galaxy-dominated and Active Galactic Nuclei (AGN)-dominated models from images in the presence of realistic noise is presented in \citet{Hassan2017}.
\par In this work, we employ a neural network, to be more specific a Progressively Growing Generative Adversarial Network (PGGAN) (that we name \textsc{21cmGAN}), for the \emph{creation} of two-dimensional tomographic 21cm signals, conditioned on a three-dimensional parameter vector.
Our implementation of \textsc{21cmGAN} in Tensorflow \citep{Abadi2016}, as well as the trained neural network are publicly available \footnote{\url{https://github.com/FloList/21cmGAN}}. Different input formats for the tomography samples are supported, and adapting the code to other 21cm catalogues that probe different parameter spaces should be possible without major difficulties.

\par This paper is structured as follows: in Section \ref{sec:21cm}, we give a bird's eye view of the evolution of the 21cm signal between Recombination and Reionisation. In Section \ref{ref:dataset}, we introduce the public data set from \citet[][henceforth \citetalias{Semelin2017}]{Semelin2017}, which we use as training data for our neural network. Then, we turn towards generative models in Section \ref{sec:PGGANs} and summarise the major concepts behind (PG)GANs. In Section \ref{sec:results}, we present our results and assess the quality of the generated samples by analysing the global 21cm signal, the power spectrum at fixed redshift, and the pixel distribution function (PDF). In Section \ref{sec:ABC}, we show how parameter inference can be done using the trained PGGAN. To this end, we consider two methods that use different parts of the neural network (\emph{generator} and \emph{critic}). We conclude this paper in Section \ref{sec:conclusions}. 

\section{21cm tomography}
\label{sec:21cm}
\subsection{21cm brightness temperature}
The differential brightness temperature of the 21cm line against the CMB is commonly written as (e.g. \citealt{Furlanetto2006})
\begin{equation}
\label{eq:T_b}
\begin{aligned} 
\delta T_{b} &= \frac{T_s - T_\gamma}{1 + z}\left(1 - e^{-\tau_\nu}\right) \\ 
&\approx 27 x_{\text{H}_\text{I}}(1+\delta)\left(\frac{T_{s}-T_{\gamma}}{T_{s}}\right) \left(1+\frac{1}{H(z)} \frac{\text{d} v_{\|}}{\text{d} r_{\|}}\right)^{-1} \\  
&\quad \times \left(\frac{1+z}{10}\right)^{\frac{1}{2}}\left(\frac{\Omega_{b}}{0.044} \frac{h}{0.7}\right)\left(\frac{\Omega_{m}}{0.27}\right)^{\frac{1}{2}} \text{mK}, \end{aligned}
\end{equation}
where $\tau_\nu$ is the optical depth, $T_s$ is the spin temperature of the 21cm transition, $T_\gamma$ is the background temperature of the CMB, $x_{\text{H}_\text{I}}$ is the local H$_\text{I}$ fraction, $H(z)$ is the Hubble parameter as a function of redshift $z$, $\delta$ is the local fractional overdensity, $\Omega_b$ and $\Omega_m$ denote the fractional energy content of baryonic matter and total matter, respectively, and $h$ is the dimensionless Hubble constant defined by $H_0 = h \times 100 \ \text{km} \ \text{s}^{-1} \ \text{Mpc}^{-1}$. Moreover, $\text{d} v_{\|} / \text{d} r_{\|}$ is the velocity gradient along the line of sight, which entails redshift space distortions \citep{Barkana2005}. The locality of the non-cosmological quantities in the above equation is what causes the spatial variations of $\delta T_{b}$.
\par The spin temperature $T_s$ is related to the ratio of number densities in each of the two hyperfine states $n_0$ (1s singlet) and $n_1$ (1s triplet) and is given as
\begin{equation}
    \frac{n_1}{n_0} = 3 \exp\left(-T_\star / T_s\right),
\end{equation}
where $T_\star = h \nu_{\text{21cm}} / K_B = 0.068 \ \text{K}$. As derived by \citet{Field1958}, the spin temperature $T_s$ can be written as a weighted sum of the CMB temperature $T_\gamma$, the kinetic temperature $T_\text{kin}$, and the colour temperature $T_{\alpha}$:
\begin{equation}
    T_{s} =\frac{T_{\gamma}+y_{\text{kin}} T_{\text{kin}}+y_{\alpha} T_{\alpha}}{1+y_{\text{kin}}+y_{\alpha}}.
\end{equation}
Note that an alternative convention expresses the inverse spin temperature $T_s^{-1}$ as the weighted sum of inverse temperatures. The colour temperature is tightly coupled to the kinetic temperature, i.e. $T_\alpha \approx T_\text{kin}$ \citep{Field1959}.
\par The interplay between mechanisms that drive $T_s$ towards $T_\gamma$ or $T_{\text{kin}}$, respectively, is expressed in the equation above by the coupling coefficients $y_{\text{kin}}$ (accounting for atomic collisions) and $y_{\alpha}$ (accounting for the Wouthuysen--Field effect, \citealt{Wouthuysen1952, Field1958}).
This decomposition is motivated by the following considerations (see e.g. \citealt{Furlanetto2006, Pritchard2012, Mesinger2011} for more detailed explanations): In the post-recombination Universe between $200 \lesssim z \lesssim 1100$, the IGM is collisionally coupled to the CMB via Compton scattering between CMB photons and leftover free electrons, leading to $T_s = T_{\text{kin}} = T_\gamma$ and therefore $\delta T_b = 0$. Afterwards, the IGM cools adiabatically according to $T_{\text{kin}} \propto (1 + z)^2$, faster than the CMB whose temperature scales as $T_\gamma \propto (1 + z)$. Hence $\delta T_b < 0$, and the 21cm signal can be detected for the first time in absorption. As the IGM gradually disperses due to the expansion of the Universe, collisional cooling becomes inefficient, and the spin temperature approaches again the CMB temperature, reaching equality at $z \sim 30$ where $\delta T_b \approx 0$. 
\par The next transition is brought about by the onset of the first astrophysical sources, which 
start coupling $T_s$ again to $T_{\text{kin}}$. This coupling arises from the Wouthuysen--Field effect due to the background Lyman-$\alpha$ radiation. Since $T_s \sim T_{\text{kin}} < T_\gamma$, the 21cm signal is once more in absorption, marking the advent of the Cosmic Dawn. Spatial fluctuations of the 21cm signal emerge from inhomogenities of the Lyman-$\alpha$ background, as well as from density variations. The Lyman-$\alpha$ coupling gradually intensifies until it eventually saturates. From then on, the 21cm signal is no longer susceptible to Lyman-$\alpha$ fluctuations. X-ray heating from astrophysical sources leads to growing temperature differences between hot, overdense regions and the cold bulk of the gas, and the 21cm signal from heated regions can be seen for the first time in emission. Gas temperature fluctuations now contribute to the brightness temperature fluctuations. \par As the X-ray heating continues and becomes dominant over cosmic expansion in the entire IGM, $T_{\text{kin}}$ surpasses $T_\gamma$ and keeps rising, eventually giving $T_s \sim T_{\text{kin}} \gg T_\gamma$. 
The 21cm line is globally in emission, and the \orange{contribution of the} spin temperature in Equation \eqref{eq:T_b} becomes negligible.
As reionisation proceeds and an increasing number of regions \orange{become} ionised, fluctuations are sourced by spatially varying ionisation fractions. 
With the end of reionisation, the 21cm signal essentially disappears (see e.g. \citealt{Wyithe2008} for the post-reionisation signal from damped Lyman-$\alpha$ systems). Note however that the timing, duration, and order of the transitions involving astrophysical sources are uncertain and model-dependent \citep{Pritchard2007}.

\section{The 21SSD data set}
\label{ref:dataset}
We use the publicly available data set from \citetalias{Semelin2017} that contains lightcones of the 21cm brightness temperature during the EoR in a redshift range of $z = 15$ to $6$ for a three-dimensional parameter space. In what follows, we briefly summarise the fixed and varying parameters of the simulations, but the interested reader is referred to said paper for a detailed discussion. 
\subsection{Overview of the simulations}
\par The lightcones have been produced from fully coupled radiative hydrodynamic simulations in a box of side length $200 \ h^{-1} \ \text{Mpc}$ with $1024^3$ particles, run with the parallel Tree+SPH code \textsc{Licorice} \citep{Semelin2002, Semelin2007, Baek2009, Baek2010, Vonlanthen2011, Semelin2016}. Monte-Carlo ray-tracing on an adaptive mesh is used for the computation of X-ray and UV radiative transfer, and both hard and soft X-ray emission is enabled. The only feedback from star formation is photo-heating. The Lyman-$\alpha$ flux is calculated in a post-processing step on a fixed grid consisting of $512^3$ cells, emitting $4 \times 10^8$ photons once every $10^7 \ \text{yr}$. \citet{Ade2016} cosmology is assumed, and the simulations start at $z = 100$. 
\par Several astrophysical parameters have been fixed such that the simulation results comply with observational constraints (see \citealt{Bouwens2015, Bouwens2015a} and references therein): a Kennicutt--Schmidt law with exponent 1 and $c_\text{eff} = 2 \ \text{Gyr}^{-1}$ models the star formation in regions with comoving overdensity $> 100$, i.e. $\orange{\text{d}\rho_s / \text{d}t} = \rho_g / (2 \ \text{Gyr})$, and a Salpeter initial mass function (IMF) is adopted for the star population, truncated below $1.6$ and above $120 \ \text{M}_\odot$, respectively. The escape fraction of ionising UV photons from unresolved structures is set to $0.2$.

\subsection{The three-dimensional parameter space} 
The \citetalias{Semelin2017} data set explores a three-dimensional parameter space, where the varying parameters are the X-ray emissivity $f_X$, the hard-to-soft ratio of the X-ray emission $r_{h/s}$, and Lyman band emissivity $f_\alpha$.
\begin{itemize}[leftmargin=.4cm]
    \item $f_X$:
 After the mean kinetic temperature of the neutral gas has decreased monotonously since recombination, X-ray heating triggers a turnaround and eventually causes $T_s = T_\text{kin} > T_\gamma$. Assuming that the local proportionality between X-ray luminosity and star formation rate (SFR) holds for high redshifts, one can parametrise \citep{Furlanetto2006}
 \begin{equation}
     L_X = 3.4 \times 10^{40} f_X \left(\frac{\text{SFR}}{1 \ \text{M}_{\odot} \ \text{yr}^{-1}}\right) \ \text{erg} \ \text{s}^{-1},
 \end{equation}
with the unknown renormalisation parameter $f_X$ accounting for the extrapolation to high redshifts. 

    \item $r_{h/s}$:
In the simulations, hard X-rays from X-ray binaries receive a special treatment due to their large mean free path, which renders the tracking of all hard X-ray photons over the course of the entire simulation unfeasible (see \citetalias{Semelin2017} for details). The description of their spectral properties follows \citet{Fragos2013}. In contrast, AGNs produce soft X-rays between $0.1$ and $2 \ \text{keV}$, with a spectral index of $1.6$. Soft X-rays are believed to carry most of the energy \citep{Furlanetto2006}, but their shorter mean free paths leads to a more localised energy deposition which generates spatial fluctuations. The fraction of hard X-rays $r_{h/s}$ is defined as 
\begin{equation}
    r_{h/s} = \frac{f_X^{\text{XRB}}}{f_X^{\text{XRB}} + f_X^{\text{AGN}}},
\end{equation}
where $f_X^{\text{XRB}}$ and $f_X^{\text{AGN}}$ are the X-ray emissivities of X-ray binaries and AGNs, respectively. 
   
   \item $f_\alpha$:
In the \citetalias{Semelin2017} data set, the efficiency of the Lyman band emission is described by one free parameter $f_\alpha$. The energy emission from a stellar population within a frequency band $[\nu_1, \nu_2]$ is given by
\begin{equation}
    E(\nu_1, \nu_2) = \int_{\nu_1}^{\nu_2} \int_{M_\text{min}}^{M_\text{max}} \xi(M) \, L(M, \nu) \, T_\text{life}(M) \ \text{d}M \, \text{d}\nu,
\end{equation}
where $M$ is stellar mass, $M_\text{min} = 1.6 \ \text{M}_\odot$, $M_\text{max} = 120 \ \text{M}_\odot$, $\xi(M)$ is the IMF, $T_\text{life}(M)$ is the lifetime of a star as a function of its mass, $L(M, \nu)$ is the energy emission per time and per frequency from a star of mass $M$ at frequency $\nu$. The parameter $f_\alpha$ is then defined as the proportionality factor between the energy as produced in the simulation $E_{\text{eff}}$ and the theoretical energy emission in the relevant frequency band, that is
\begin{equation}
    E_{\text{eff}}(\nu_\alpha, \nu_\text{limit}) = f_\alpha \ E(\nu_\alpha, \nu_\text{limit}),
\end{equation}
with the Lyman-$\alpha$ frequency $\nu_\alpha$ and Lyman limit frequency $\nu_\text{limit}$. Higher Lyman band emission efficiencies $f_\alpha$ produce more prominent troughs and peaks in the differential brightness temperature.

\end{itemize}
\subsection{Data creation for the PGGAN training}
The \citetalias{Semelin2017} data set contains simulations for each of the possible combinations of the following parameter values:
\begin{equation*}
    \begin{aligned}
    \hspace{0.4cm}& f_X: & \quad 0.1,& & \pquad 0.3,& & \pquad 1,& & \pquad 3,& & \pquad 10, \\
    \hspace{0.4cm}& r_{h/s}: & \quad 0,& & \pquad 0.5,& & \pquad 1,& & & &\\
    \hspace{0.4cm}& f_\alpha: & \quad 0.5,& & \pquad 1,& & \pquad 2,& & & &
    \end{aligned}
\end{equation*}
thus 45 simulations in total. For each simulation, three high-resolution lightcones of the differential brightness temperature are available, corresponding to observers aligned with the three coordinate axes, giving a total of 135 lightcones. The line-of-sight dimension of 21cm images is special as it represents measurements at different frequencies, from which the associated redshifts and hence comoving distances can be inferred to reconstruct a 3D distribution. The frequency resolution is typically chosen such that the resulting spatial line-of-sight resolution matches the angular one. This is also the case for the \citetalias{Semelin2017} lightcones whose cells have an equal $\Delta \nu$ that leads to almost cubic cells (of slightly varying comoving thickness). Although brightness \orange{temperature} lightcones at $\sim 3'$ ($32 \times 32 \times 256$ pixels, with the line-of-sight dimension being the longest) and $\sim 6'$ ($16 \times 16 \times 128$ pixels) resolution are provided in the public data set for convenience, we start from the high-resolution lightcones ($1024 \times 1024 \times 8096$ pixels) in order to obtain more slices as a large data set size is crucial for deep learning. We cut \orange{$2048$ slices from each lightcone by fixing pixels in the first and second dimension, respectively,} yielding in total $276480 = 2 \times 1024 \times 135$ 2D tomography samples ($6144$ samples for each set of parameters) at a resolution of $1024 \times 8096$. The shorter (vertical) dimension of the 2D samples is perpendicular to the line of sight and covers a comoving length of $200 \ h^{-1} \ \text{Mpc}$, whereas the longer (horizontal) dimension is aligned with the line of sight and comprises a redshift range of $z = 15$ to $6$ (from left to right). If \textsc{21cmGAN} is to be applied to smaller data sets, one might resort to data augmentation methods such as mirroring. We subsequently downscale the samples by averaging over adjacent pixels to obtain a resolution of $32 \times 256$ pixels. From these samples, we create versions from resolution $1 \times 8$ to $32 \times 256$, corresponding to each stage of the PGGAN training (see Subsection \ref{subsec:growing}).

\section{PGGANs}
\label{sec:PGGANs}
The two most prominent representatives within the class of generative models are Variational Auto-Encoders \citep[VAEs,][]{Kingma2013} and Generative Adversarial Networks \citep[GANs,][]{Goodfellow2014}. While GANs are notoriously hard to train, they are more flexible and generally produce images of higher quality \citep[e.g.][however, see \citealt{Razavi2019} for recent progress on VAEs]{Arora2017a}, which is why we choose them for this work.
\subsection{GANs}
The setting of a standard GAN in our context is the following: two players, the \emph{generator} $G$ and the \emph{discriminator} $D$, compete in a min-max-game against each other. Let $p(\mathbf{x} | \mathbf{\theta})$ be the probability distribution function of the differential brightness temperature images $\mathbf{x} \in [-1, 1]^{H \times W}$ (where $H$ and $W$ stand for the height and width of the images)\footnote{The pixel values can also be scaled to other intervals, e.g. $[0, 1]$ instead of $[-1, 1]$.}, given a vector of reionisation parameters $\mathbf{\theta}$.  The \emph{generator} outputs samples from a distribution $p_G(\mathbf{x} | \mathbf{\theta})$ and attempts to shift $p_G$ towards $p$ as the training progresses. On the other hand, the \emph{discriminator} tries to distinguish the images drawn from $p_G$ from the true samples taken from $p$. Mathematically, the \emph{generator} and \emph{discriminator} are mappings $G : (\mathbf{z}; \mathbf{\theta}) \mapsto G(\mathbf{z} | \mathbf{\theta})$ and $D : (\mathbf{x}; \mathbf{\theta}) \mapsto D(\mathbf{x} | \mathbf{\theta}) \in [0, 1]$. The random noise vector $\mathbf{z}$, which is usually drawn from a standard uniform or normal distribution $p_\mathbf{z}$, introduces randomness and enables the \emph{generator} to create a genuine distribution of samples for each $\mathbf{\theta}$, rather than a deterministic mapping. The \emph{discriminator} input $\mathbf{x}$ is either a real sample, i.e. $\mathbf{x} \sim p(\mathbf{x} | \mathbf{\theta})$, or has been \orange{produced} by the \emph{generator}, i.e. $\mathbf{x} \sim p_G(\mathbf{x} | \mathbf{\theta})$, and the \emph{discriminator} output estimates the probability of the former. As the probability distributions in our case are conditioned on the reionisation parameters $\mathbf{\theta}$, this variant of GANs is known as a conditional GAN \citep{Mirza2014}. The training objective is thus tantamount to finding a saddle point 
\begin{equation}
(G^*, D^*) = \text{arg} \, \min_G \max_D \mathcal{L}_\text{GAN}(G, D; \mathbf{\theta}),   
\end{equation}
where the \orange{function} $\mathcal{L}_\text{GAN}$ is defined as
\begin{equation}
\begin{aligned}
\mathcal{L}_\text{GAN}(G, D; \mathbf{\theta}) &= \mathbb{E}_{\mathbf{x} \sim p(\mathbf{x})} \left[\log D(\mathbf{x}|\mathbf{\theta})\right] \\
&\quad+ \mathbb{E}_{\mathbf{z} \sim p_\mathbf{z}(\mathbf{z})} \left[\log\left(1 - D(G(\mathbf{z}|\mathbf{\theta})|\mathbf{\theta})\right)\right].
\end{aligned}
\end{equation}
While the mathematical theory of such problems is well established (e.g. \citealt{nash1950equilibrium}), the practical training of GANs is intricate and often unstable \citep{Arjovsky2017a}. A problem commonly encountered is that of so-called mode collapse, which describes the situation when the \emph{generator} output is but a small subset of the real data distribution $p$. Whereas the \orange{loss} $\mathcal{L}_\text{GAN}$ is based upon the Jensen--Shannon divergence, different \orange{loss functions} have been proposed that rely on other distance measures such as $f$-divergences \citep{Nowozin2016}. An interesting approach was introduced by \citet{Arjovsky2017}, who advocate the use of the Wasserstein distance (also known as Earth Mover's distance) that provides meaningful gradients even when $p_G$ and $p$ are far apart from each other. The $q$-Wasserstein distance $W_q(p_1, p_2)$ between two probability distributions $p_1$ and $p_2$ is defined as 
\begin{equation}
\label{eq:Wasserstein_dist}
    W_q(p_1, p_2) = \left( \inf_{\gamma \in \Pi(p_1, p_2)} \mathbb{E}_{(a, b) \sim \gamma} \left[\| a - b \|^q \right] \right)^{\frac{1}{q}},
\end{equation}
where $\Pi(p_1, p_2)$ denotes the set of all joint distributions $\gamma(a, b)$ with marginals $p_1$ and $p_2$, respectively, and $q \geq 1$. In the case of $q = 1$, the Wasserstein distance gives the minimum amount of work needed for moving the ``probability mass'' piled up according to $p_1$ to the distribution $p_2$, where work is defined as the product of probability mass and distance by which the mass is moved. An equivalent characterisation with more practical relevance is obtained using the Kantorovich--Rubenstein duality, yielding
\begin{equation}
    W_1(p_1, p_2) = \sup_{\|f\|_L \leq 1} \left( \mathbb{E}_{a \sim p_1} \left[f(a) \right] - \mathbb{E}_{b \sim p_2} \left[f(b)\right]\right),
\end{equation}
where the supremum is taken over all 1-Lipschitz continuous functions. While this Lipschitz constraint is achieved by clipping the weights of the neural network in \citet{Arjovsky2017}, \citet{Gulrajani2017} propose to make use of the fact that a differentiable function $f$ is 1-Lipschitz if and only if its gradients are bounded by 1 everywhere. This motivates replacing the weight clipping by a penalty term, giving rise to WGAN-GPs (Wasserstein GANs with Gradient Penalty). The appropriate \orange{loss function} $\mathcal{L}_\text{WGAN-GP}$ for our saddle point problem reads as
\begin{equation}
\label{eq:WGAN-GP}
\begin{aligned}
    \mathcal{L}_\text{WGAN-GP}(G, D; \mathbf{\theta}) &=
     \mathbb{E}_{\mathbf{x} \sim p(\mathbf{x})} \left[D(\mathbf{x}|\mathbf{\theta})\right] \\ 
    &\quad- \mathbb{E}_{\mathbf{z} \sim p_\mathbf{z}(\mathbf{z})} \left[D(G(\mathbf{z}|\mathbf{\theta}) | \mathbf{\theta})\right] \\
    &\quad+ \lambda \, \mathbb{E}_{\orange{\tilde{\mathbf{x}}} \sim \orange{p_{\tilde{\mathbf{x}}}}} \left[\left(\|\orange{\nabla_{\tilde{\mathbf{x}}}} \orange{D(\tilde{\mathbf{x}} | \mathbf{\theta})} \|_2 - 1 \right)^2 \right]. 
\end{aligned}
\end{equation}
Here, $\orange{\tilde{\mathbf{x}}} = \alpha \mathbf{x} + (1-\alpha) G(\mathbf{z}|\mathbf{\theta})$ with a random number $\alpha \sim U[0,1]$, and the factor $\lambda$ determines the strength of the gradient penalty. Using the gradient penalty in lieu of the weight clipping has been shown to improve the image quality and convergence. In the context of WGANs, the \emph{discriminator} is referred to as the \emph{critic} since it no longer measures the probability of $\mathbf{x} \sim p(\mathbf{x} | \mathbf{\theta})$ but now produces a score that gauges how real $\mathbf{x}$ seems to the \emph{critic} (in particular, the output of the \emph{critic} is not confined to $[0, 1]$). The aim of the \emph{critic} is to maximise the difference between scores for real and fake samples.

\subsection{Progressive growing}
\label{subsec:growing}
In this work, we present the emulation of 21cm tomography samples consisting of $32 \times 256$ pixels. 
The corresponding angular resolution is \orange{$3.1' - 3.8'$ (for $z = 15 - 6)$}, which is within the typical range of the SKA. As an example, the Baseline Design of SKA1-Low \citep{Dewdney2013} will have a projected noise of $\sim 10 \ \text{mK}$ at a frequency of $\nu = 140 \ \text{MHz}$ (corresponding to redshift $z \sim 9$) on scales of $3'$ \citep{Mellema2015}. \orange{With increasing} resolution (and hence number of pixels per image), training GANs becomes increasingly time and memory consuming\orange{. Furthermore, the more pixels the \emph{discriminator} has at its disposal, the easier it becomes to discern real and fake images, while the task of the \emph{generator} to generate high quality output consisting of many pixels becomes harder. This can lead to an imbalance and cause the GAN training to fail.} Recently, much work has been focused on methods for the generation of high-resolution images with GANs, and applying techniques such as an intermediate latent space in the \emph{generator} \citep{Karras2018}, self-attention layers \citep{Zhang2018}, spectral normalisation \citep{Miyato2018}, or a truncation trick \citep{Brock2018} has led to remarkable results. A major breakthrough is the work by \citet[][hereafter \citetalias{Karras2017}]{Karras2017}, who progressively increase the resolution and number of layers during training, obtaining high-quality $1024 \times 1024$ images. Although our resolution of $32 \times 256$ pixels is rather modest, we use a Progressively Growing GAN (PGGAN), mainly because of two reasons: firstly, we observe a lower tendency towards mode collapse when building up our GAN progressively and secondly, our neural network architecture can easily be extended to generate higher resolution tomography samples that might be needed for future interferometry experiments. Of course, creating 3D samples at the full resolution of the \citetalias{Semelin2017} catalogue of $1024 \times 1024 \times 8096$ pixels with GANs is still a long way off.
\par During the training, new layers are gradually added to both the \emph{generator} and the \emph{critic}. This happens in a smooth manner, allowing the GAN to slowly transition to a higher resolution. Once part of the GAN, the weights of each layer remain trainable until the end of the training. While the GAN is trained on lower resolution images, downscaled images are shown to the \emph{critic}. For more details on the progressive growing of the GAN, we refer the interested reader to \citetalias{Karras2017}, whose approach we largely follow. 

\subsection{Network architecture}
\begin{figure}
  \centering
  \noindent
  \resizebox{\columnwidth}{!}{
  \includegraphics{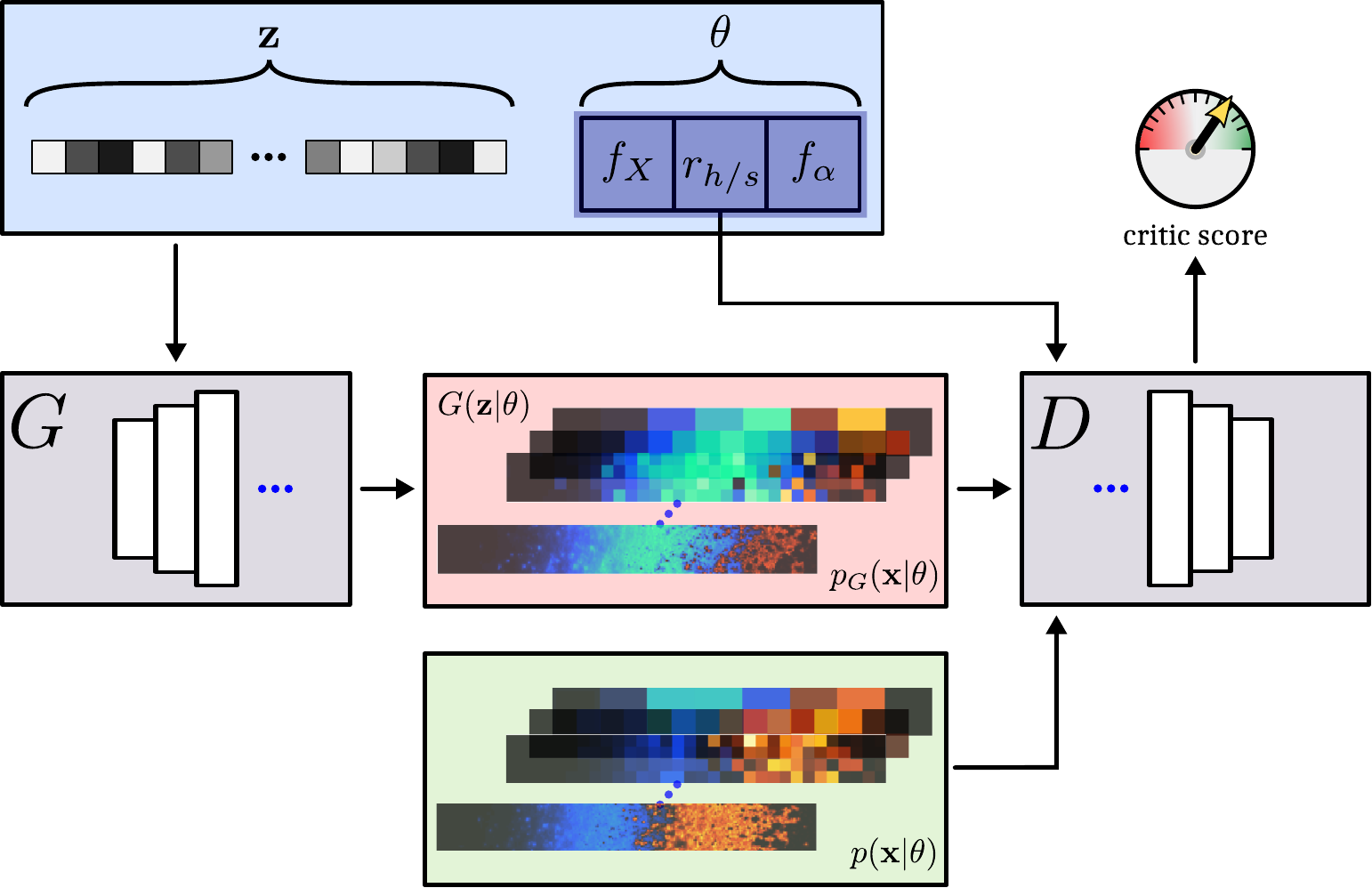}
  }
  \caption{Schematic representation of \orange{\textsc{21cmGAN}}: given a random noise vector $\mathbf{z}$ and a parameter vector $\mathbf{\theta}$, the \emph{generator} $G$ produces an image $G(\mathbf{z}|\mathbf{\theta})$ in an attempt to fool the \emph{critic} $D$. Images from the \emph{generator} and from the training data set are shown to the \emph{critic}, whose task it is to assign high scores to real samples and low scores to GAN-produced samples. Importantly, the \emph{critic} judges the image quality of the samples \emph{given the parameter vector $\mathbf{\theta}$}. Accordingly, a real sample corresponding to parameters $\mathbf{\theta}_1$ shown to the \emph{critic} together with very different parameters $\mathbf{\theta}_2$ will receive a low score from the \emph{critic}. Blue triple dots stand for growth as the training proceeds: in order to produce samples at gradually increasing resolution, new layers are smoothly faded in during the course of the training. This is done for the \emph{generator} and the \emph{critic} in a symmetric way.}
  \label{fig:PGGAN_sketch}
\end{figure}
Our network architecture borrows from \citetalias{Karras2017}, which is why we mainly describe the differences in what follows. The workflow of our PGGAN is schematically depicted in Figure \ref{fig:PGGAN_sketch}, and the details are listed in Appendix \ref{sec:NN_details}.
\par Our \emph{generator} network is a CNN whose layers successively construct the image up to the desired resolution. We take the latent vector to be of length $512$, where we fix the first three components to be the entries of $\mathbf{\theta} = (f_X, r_{h/s}, f_\alpha)$ and append $509$ independent and identically distributed drawings from a standard normal distribution, the latter forming the vector $\mathbf{z}$.  
Thus, we condition the \emph{generator} by manually encoding information in the otherwise random latent vector. Whereas benchmark data sets for GANs are typically composed of square images, our images have an aspect ratio of $1 \ \colon 8$. For this reason, we choose $1 \times 8$ as the base resolution of our GAN. First, the latent vector is mapped to the base resolution via a fully connected layer. Then, convolutional layers and upsampling layers follow, which gradually increase the spatial resolution until the desired resolution at the respective stage of the training is reached.  
\par For the \emph{critic}, we use a structure that is essentially symmetric with respect to the \emph{generator} architecture. Since the input for the \emph{critic} is an image $\mathbf{x} \in [-1, 1]^{H \times W}$, we cannot simply append the three-dimensional parameter vector to the input as we do for the \emph{generator}. Instead, we concatenate the three parameters as constant additional image channels. \orange{We leave the implementation of more sophisticated conditioning methods for the \emph{critic} as proposed by \citet{Miyato2018a} for future work.} While the \emph{critic} determines autonomously which features are relevant for assessing the image quality, there are two major requirements for the generated samples that we point out: firstly, the global brightness temperature, that is the vertical average in the 2D case considered herein, should agree with the samples from the \citetalias{Semelin2017} catalogue. Secondly, the fluctuations of the mock images should exhibit the same statistics as their \citetalias{Semelin2017} counterparts. We encourage the \emph{critic} to evaluate these properties by appending the vertically averaged brightness temperature $\langle \delta T_b \rangle$ and the difference towards it, i.e. $\delta T_b - \langle \delta T_b \rangle$, as additional image channels. Altogether, the input for the \emph{critic} hence takes the shape $6 \times 32 \times 512$ (1 image channel, 2 extra channels, and 3 parameter channels).

\subsection{Training}
We train our PGGAN on 4 Nvidia Tesla Pascal P100 GPUs on the supercomputer Raijin, which is located in Canberra and is part of the Australian National Computational Infrastructure (NCI). We use a batch size of 8 per GPU and execute 40000 iterations for each growth stage of the PGGAN. The training alternates between transitional stages during which the next higher resolution (with twice as many pixels in each dimension) is faded in, and stabilising stages, where the PGGAN is trained at the respective resolution. Starting at a resolution of $1 \times 8$, this means that the training consists of $6$ stabilising stages, separated by $5$ transitional stages, to arrive at resolution $32 \times 256$. This amounts to $440000$ iterations in total, requiring $\sim 64 \ \text{h}$ of wall time. We take the WGAN-GP loss function as defined in Equation \eqref{eq:WGAN-GP} with $\lambda = 10$, plus a small additional penalty term $-0.001 \, \mathbb{E}_{\mathbf{x} \sim p(\mathbf{x})} [D(\mathbf{x} | \mathbf{\theta})^2]$ in order to keep the \emph{critic} scores close to zero and prevent them from diverging. The minimisation is done using an Adam optimiser \citep{Kingma2014}. 

\subsection{Practical usage}
We briefly elaborate on how \textsc{21cmGAN} can be utilised in practice for the generation of 21cm tomography samples and for aiding parameter inference. The first and simplest option is to use the trained PGGAN presented in this work, which readily produces tomography samples at a resolution of $\gtrsim 3'$. If required, instrument-specific noise and foregrounds can be added to the generated images. Another possibility is to re-train (a modified version of) \textsc{21cmGAN} on a different set of tomography images. First, a parameter space needs to be defined that describes the processes impacting the 21cm signal. In this work, X-ray emissivity, ratio of hard to soft X-rays, and Lyman-$\alpha$ efficiency are considered; however, alternative parameter spaces are spanned for instance by the ionising efficiency of high-redshift galaxies $\zeta$ (which itself depends on the escape fraction of ionising UV photons, the fraction of galactic gas in stars, the number of times a hydrogen atom recombines, and the number of ionising photons that baryons in stars produce), the mean free path of ionising photons in ionised regions, and the minimum virial temperature for star formation in haloes \citep{Greig2015}. Furthermore, different models for star formation and supernova feedback can be considered \citep{Mesinger2016}. Once the parameter space is set, a suitable sampling strategy is required, e.g. grid-based sampling, Latin Hypercube Sampling, or an adaptive metric-based approach as proposed by \citet{Eames2019}. Then, 
21cm samples are generated for the chosen parameter sets, either using a hydrodynamic simulation with radiative transfer or an approximate method. The produced samples (or 2D slices thereof) serve as training data for \textsc{21cmGAN}, which learns the dependence of the 21cm tomography samples on the individual parameters. The trained neural network can create output samples for arbitrary points in parameter space (evidently, caution is needed when the neural network is evaluated on parameters that lie outside the range in parameter space on which the neural network was trained, or if the training data stems from an extremely coarse sampling of the parameter space). 
\par The trained neural network represents a sampler from the probability distribution function $p_G(\mathbf{x}|\mathbf{\theta}) \approx p(\mathbf{x}|\mathbf{\theta})$, which, in contrast to $p$, can be evaluated on a continuous subset in parameter space. This sampler can be utilised for backward modelling, that is parameter inference from a given tomography sample, as well: since the approximate likelihood function $\mathcal{L}_G(\mathbf{\theta}) = p_G(\mathbf{x}|\mathbf{\theta})$ is analytically intractable, methods that rely on an expression for the likelihood such as MCMC cannot be applied, however, likelihood-free methods such as Approximate Bayesian Computation (ABC) provide a suitable framework not only for finding the maximum-likelihood estimate for $\theta$, but also for obtaining error estimates. This will be demonstrated in Section \ref{sec:ABC}.

\section{Results}
\label{sec:results}
\subsection{Qualitative comparison}
\begin{figure*}
  \centering
  \noindent
  \resizebox{\textwidth}{!}{
  \includegraphics{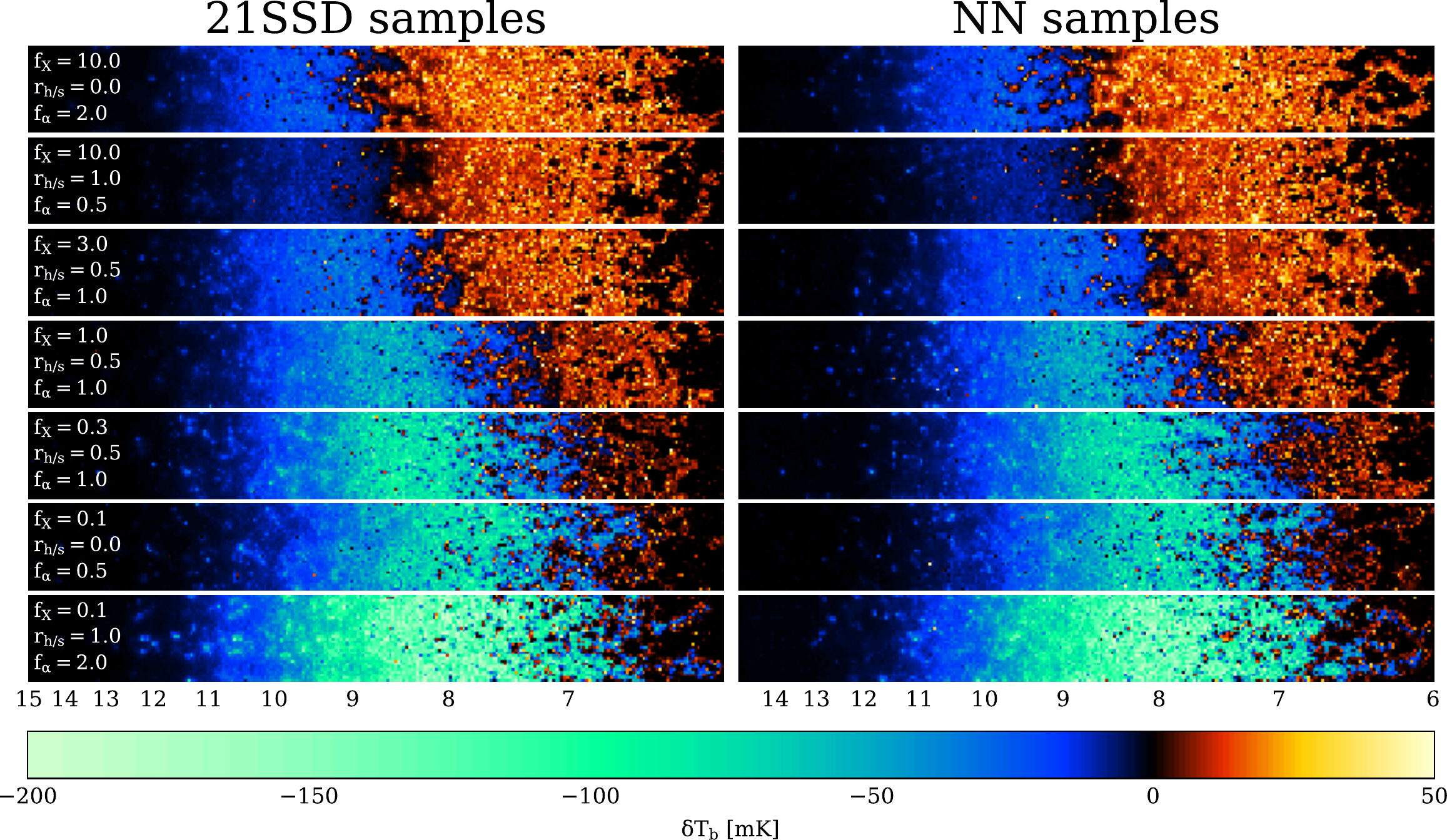}
  }
  \caption{Random tomography samples for different parameter \orange{vectors}, from redshift $z = 15$ to $z = 6$ (as indicated below the samples). The X-ray emissivity $f_X$ decreases from top to bottom. For the extreme cases $f_X = 0.1$ and $10$, two different combinations of $r_{h/s}$ and $f_\alpha$ are shown.}
  \label{fig:samples}
\end{figure*}
We start with a visual comparison between the samples generated by \textsc{21cmGAN} and those from the \citetalias{Semelin2017} catalogue. Figure \ref{fig:samples} shows random samples for different parameter sets, sorted by the strength of the emission in decreasing order (cf. Fig. 7 in \citetalias{Semelin2017} for samples at the full resolution of the catalogue). Evidently, the PGGAN has learned to map the input parameters to the correct redshifts and amplitudes of the absorption and emission regions. While the 21cm line is in emission almost everywhere at $z \sim 7 - 8$ for $f_X = 10$, the emission region for $f_X = 0.1 - 0.3$ consists of many jagged patches, which is well reproduced by the PGGAN. Also, the isolated bubbles in absorption at $z = 11 - 13$ are captured correctly. The spatial distribution of emission and absorption regions is diverse, and we do not observe any indication of mode collapse. Altogether, the visual impression of the GAN-generated tomography samples is satisfactory, and it is difficult to distinguish them from their \citetalias{Semelin2017} counterparts by eye. More random samples for selected parameter vectors in Appendix \ref{sec:samples} demonstrate the diversity of the PGGAN output and their resemblance to the \citetalias{Semelin2017} slices.

\subsection{The global 21cm signal}
\begin{figure*}
  \centering
  \noindent
  \resizebox{\textwidth}{!}{
  \includegraphics{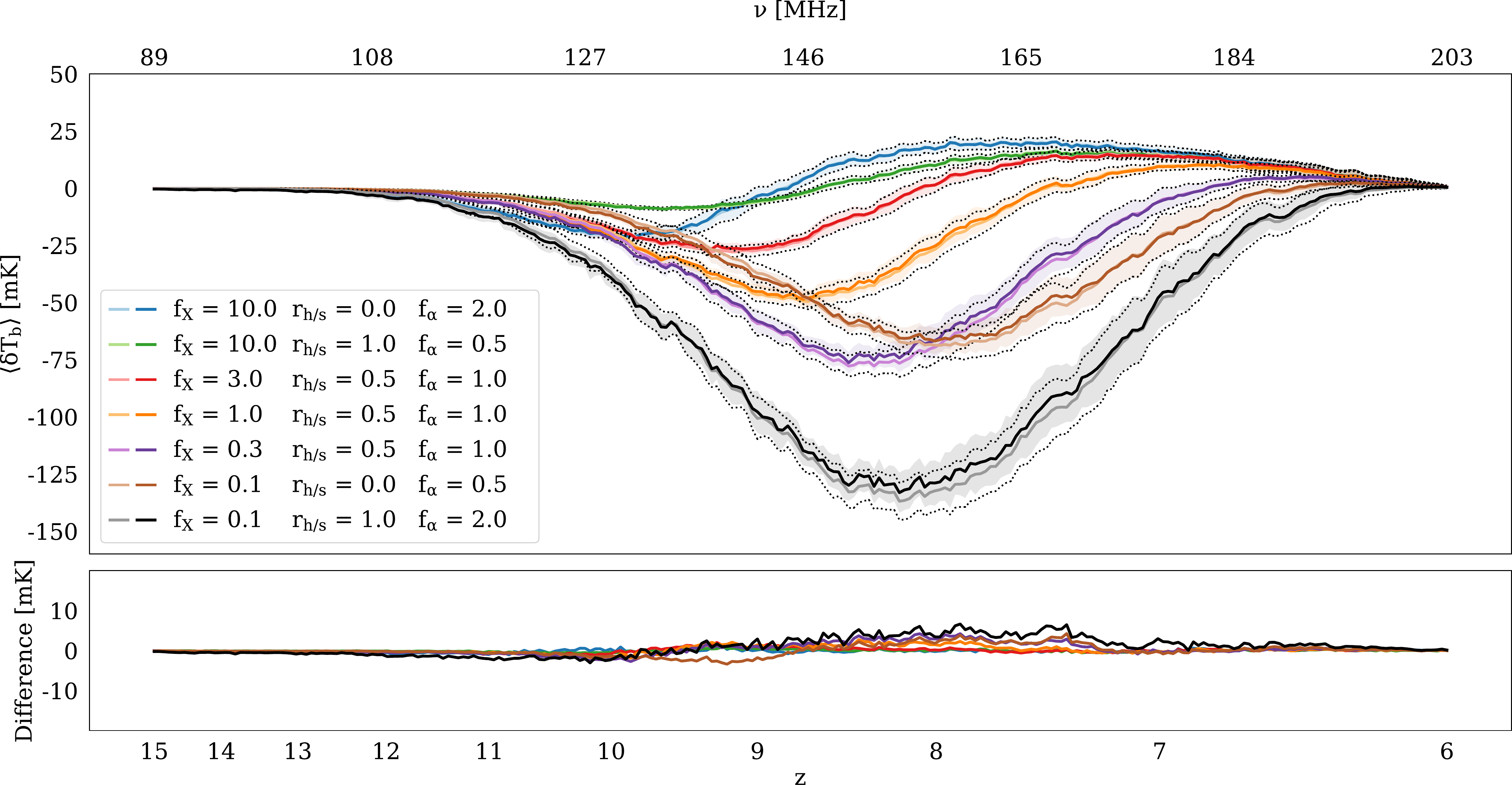}
  }
  \caption{The global 21cm signal, obtained by averaging over the vertical dimension of the samples and averaging over 6144 samples. Colours in the first / second column of the legend correspond to \citetalias{Semelin2017} / PGGAN samples, respectively. The shaded regions show the standard deviation of the brightness temperature for the PGGAN samples at each redshift, whereas the dotted lines correspond to the standard deviation for the \citetalias{Semelin2017} samples. The difference between the averages $\langle \delta T_b \rangle_{\text{PGGAN}} - \langle \delta T_b \rangle_{\text{21SSD}}$ is plotted below.}
  \label{fig:global_signal}
\end{figure*}
Figure \ref{fig:global_signal} shows the global 21cm signal $\langle \delta T_b \rangle$ as a function of redshift (solid lines, first legend column for the \citetalias{Semelin2017} samples, second column for the PGGAN samples), computed as the average over all pixel values of 6144 samples at each redshift. The 1$\sigma$-region is shaded for PGGAN samples and delimited by dotted lines for the \citetalias{Semelin2017} samples. As already seen above, the global signal is highly sensitive to $f_X$. Additionally, higher Lyman-$\alpha$ emissivities $f_\alpha$ cause earlier and more prominent peaks and troughs. The parameter $r_{h/s}$ has an effect on the average kinetic temperature of the H$_\text{I}$ gas (see Fig. 5 in \citetalias{Semelin2017}). We consider the same parameter sets as in Figure \ref{fig:samples}. The global signal of the samples from the PGGAN agrees well with that from the \citetalias{Semelin2017} catalogue for all parameter sets, and the maximum difference amounts to less than $10 \ \text{mK}$ (and much less for models with high X-ray emissivity). Also, the spread in the brightness temperature at each redshift is well reproduced, with low X-ray models exhibiting a larger spatial variability than high X-ray models. The fluctuations of the PGGAN brightness temperature follow those of the \citetalias{Semelin2017} samples, which suggests that extending the set of training images such that the resulting global signal is smoother would lead to a smoother global signal of the \citetalias{Semelin2017} samples as well.

\subsection{Power spectrum}
\begin{figure}
  \centering
  \noindent
  \resizebox{\columnwidth}{!}{
  \includegraphics{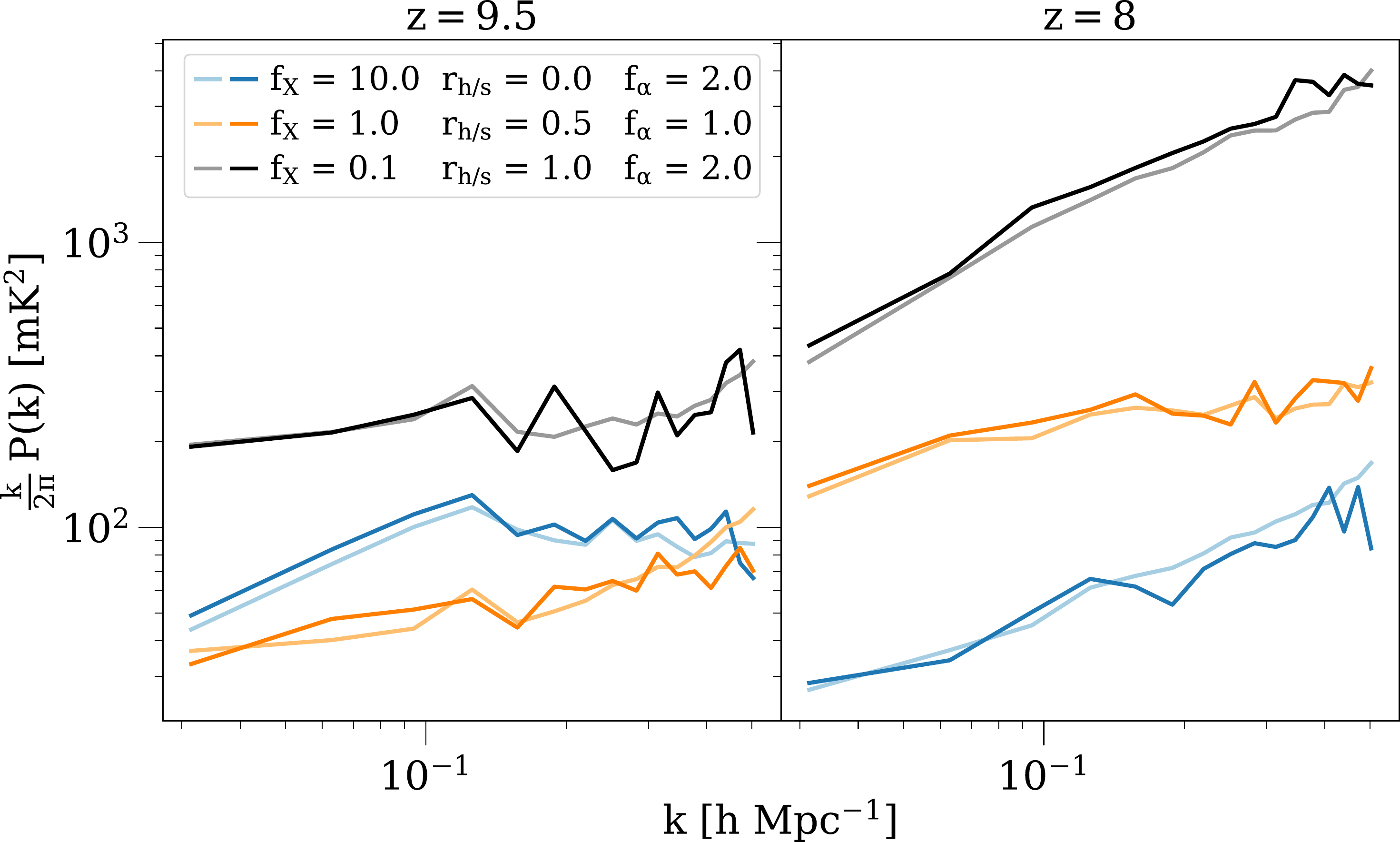}
  }
  \caption{1D power spectrum of the brightness temperature for three different models at redshifts $z = 9.5$ and $z = 8$, averaged over 6144 samples (first legend row: \citetalias{Semelin2017}, second legend row: PGGAN).}
  \label{fig:PS}
\end{figure}
A standard statistic for measuring the strength of fluctuations is the power spectrum. Figure \ref{fig:PS} shows the 1D power spectrum for two models bracketing the investigated range of $f_X$ and one model with an intermediate value of $f_X = 1$. The model with $f_X = 0.1$ has the highest power on all scales at redshifts $z = 8$ and $z = 9.5$. At $z = 9.5$, the intermediate model has the lowest power, while the power increases by $z = 8.5$. In contrast, the power of the model with $f_X = 10$ decreases on large scales as the 21cm line changes from absorption (at $z = 9.5$) to emission (at $z = 8$). The GAN-generated samples possess a similar power spectrum to the \citetalias{Semelin2017} samples at both redshifts and capture the effects of the different parameters, although the spectra are a bit noisier on small scales.

\subsection{Non-Gaussianities}
\begin{figure}
  \flushright
  \noindent
  \resizebox{\columnwidth}{!}{
  \includegraphics{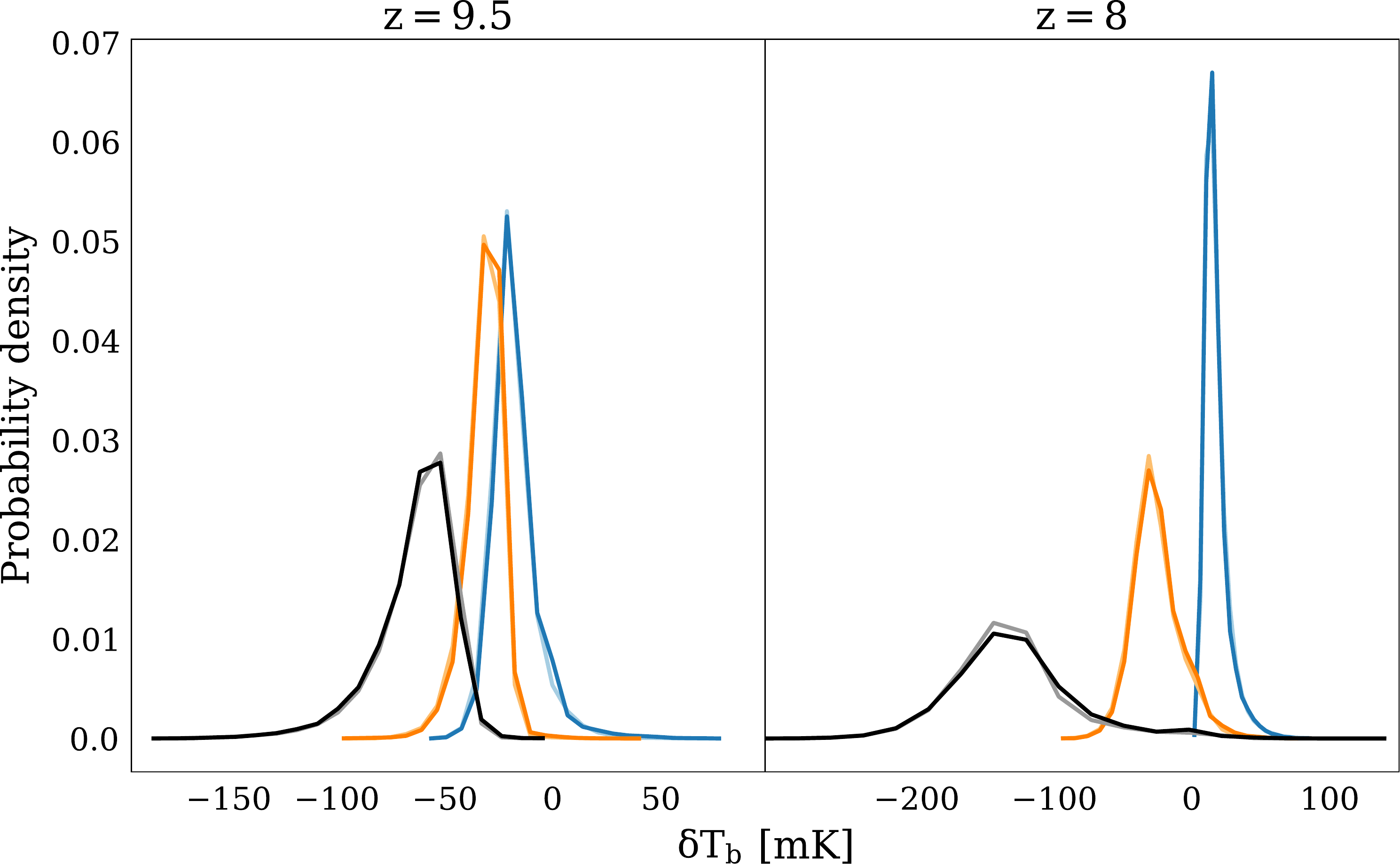}
  }
  \caption{PDF for three different models at redshifts $z = 9.5$ and $8$, obtained by binning the values of $\delta T_b$ from 6144 samples at each redshift into temperature bins, where the bin width has been taken to be $1 \ \text{mK}$. Colours have the same meaning as in Figure \ref{fig:PS}.}
  \label{fig:PDF}
\end{figure}

Due to the strong non-Gaussianity of the 21cm signal, the power spectrum alone is not sufficient for a complete description of the fluctuations. In the context of our PGGAN tomography emulator, it is therefore important that not only the power spectra match, but also the pixel distribution functions (PDFs) at each redshift. The PDF gives the distribution of the occurring brightness temperatures at each redshift. Figure \ref{fig:PDF} shows the PDF for the same models and redshifts as in Figure \ref{fig:PS}. The PDF of the model with the lowest X-ray emissivity has the widest temperature spread, which is clearly non-Gaussian but negatively skewed at $z = 9.5$. In contrast, the model with $f_X = 10$ has a narrow distribution that is positively skewed at $z = 8$. The PDFs from the GAN-generated samples closely resemble those from their \citetalias{Semelin2017} brethren; in particular, the width, height, and skewness are in excellent agreement. 
\par Alternative ways of characterising the non-Gaussian structure of the 21cm fluctuations are higher order statistics such as the bispectrum or topological measures, for instance Minkowski functionals, which are not in the scope of this work.

\subsection{Interpolating in parameter space}
\begin{figure*}
  \centering
  \noindent
  \resizebox{0.66\textwidth}{!}{
  \includegraphics{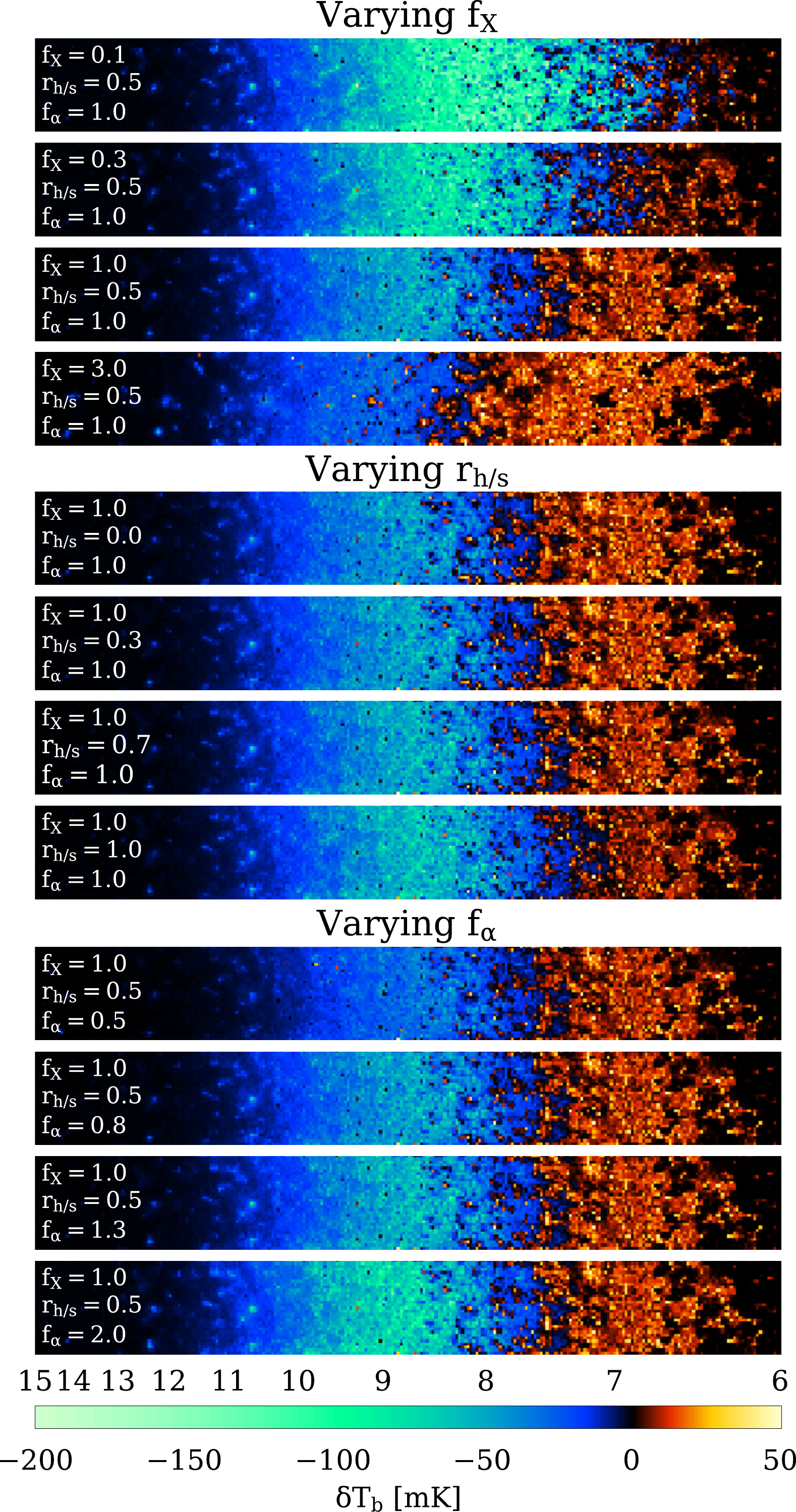}
  }
  \caption{Interpolation along the three axes of the parameter space: the parameters $f_X$, $r_{h/s}$, and $f_\alpha$ are monotonically varied (from top to bottom), while the random noise vector $\mathbf{z}$ and the other two parameters are kept fixed. Some of the selected parameter vectors were not contained in the discrete parameter space used for training, showing that the PGGAN has learned to interpolate to a continuous parameter space.}
  \label{fig:interpolation}
\end{figure*}
Keeping the 509 randomly drawn components of the noise vector $\mathbf{z}$ fixed, one can investigate how a particular sample changes as either of the reionisation parameters is varied. Figure \ref{fig:interpolation} shows an exemplary interpolation in parameter space of a random sample. The three blocks of four samples correspond to varying $f_X$, $r_{h/s}$, and $f_\alpha$ (from top to bottom), while the remaining two other parameters are kept fixed. The PGGAN was only trained on some of the depicted parameter \orange{vectors} and has learned a continuous mapping from the parameters to the samples, rather than just memorising samples that it has seen during the training. The random noise $\mathbf{z}$ governs the localisation of the stochastic brightness temperature fluctuations. 

\section{GAN-aided inference using Approximate Bayesian Computation}
\label{sec:ABC}
The trained PGGAN has learned a probabilistic mapping from the parameter space to realistic 2D tomographic samples. We now present two approaches for harnessing the PGGAN for the inverse problem, namely for inferring parameters from a tomographic sample. \orange{First, we briefly discuss uncertainty in the context of parameter inference from 21cm images.}

\orange{
\subsection{Sources of uncertainty}
In uncertainty quantification, one commonly distinguishes between two different types of uncertainty: aleatoric (statistical) uncertainty from the data and epistemic (systematic) uncertainty from the modelling. For the application of estimating astrophysical parameters from spatially resolved 21cm measurements from the EoR, aleatoric uncertainty comes from thermal noise and from sample variance. The latter is due to the individuality of the spatial fluctuations in each finite volume under consideration and is also present for the 3D 21cm lightcones in the \citetalias{Semelin2017} catalogue, each of which possesses slightly different statistics such as the 2D power spectrum. By going from 3D to 2D tomographic samples, the sample variance increases significantly since the information content per sample is reduced. When being trained, the GAN learns to generate samples whose particular spatial fluctuations are determined by the random noise vector $\mathbf{z}$, in such a way that the sample-to-sample variance agrees with that of the training data. Conversely, accurate inference of parameters from tomographic samples is limited by the sample variance. This being said, even for upcoming observations with the SKA, thermal noise will be a greater source of stochasticity as compared to sample variance \citep{Koopmans2015}. None the less, we focus in this section on aleatoric uncertainty stemming from sample variance as this is the stochasticity that the GAN has learned to generate in the training process. The extension to additional uncertainty due to thermal noise is discussed in Subsection \ref{subsec:noise}. 
\par Since aleatoric uncertainty is inherent to the tomography samples, it does not decrease as the training of the GAN proceeds. This is in contrast to the epistemic uncertainty, which describes the ignorance about the correct model that explains the data. For neural networks, it is determined by the spread of the posterior distributions of the neural network weights. In the course of the GAN training, the epistemic uncertainty should gradually decline as the distribution produced by the GAN approaches the true distribution, i.e. $p_G(\mathbf{x}|\mathbf{\theta}) \approx p(\mathbf{x}|\mathbf{\theta})$. The estimation of the epistemic uncertainty is out of the scope of this work, and we simply take the credible regions obtained for the parameters at face value.
}

\subsection{ABC rejection sampling}
A popular class of techniques that allow parameter inference given a forward simulator (the PGGAN in our case) is known as Approximate Bayesian Computation (ABC). Unlike MCMC-type methods, ABC is a likelihood-free way of inverse modelling, meaning that an analytic expression of $\mathcal{L}(\mathbf{\theta}) = p(\mathbf{x} | \mathbf{\theta})$ is not required. For simplicity, we use ABC rejection sampling, which is the easiest representative of ABC methods, but more sophisticated schemes such as ABC Sequential Monte Carlo (ABC-SMC) or Bayesian Optimization for Likelihood-Free Inference \citep[BOLFI,][]{Gutmann2015} can be applied without difficulties. 
\par Let $\hat{\mathbf{x}}$ be a tomographic sample belonging to a true unknown parameter vector $\hat{\mathbf{\theta}}$. Furthermore, let $p(\mathbf{\theta})$ be the prior distribution of the parameters, and let $\Delta_i = \Delta(\hat{\mathbf{x}}, \mathbf{x}_i)$ be a discrepancy measure between the observed sample $\hat{\mathbf{x}}$ and a generated sample $\mathbf{x}_i$. The underlying idea of ABC rejection sampling is simple: generate samples $\mathbf{x}_i$ for random parameter vectors $\mathbf{\theta}_i \sim p(\mathbf{\theta})$ and compare them to $\hat{\mathbf{x}}$. If a generated sample $\mathbf{x}_i$ is similar to $\hat{\mathbf{x}}$, i.e. $\Delta_i$ is sufficiently small, the parameters $\mathbf{\theta}_i$ should be similar to $\hat{\mathbf{\theta}}$. 

\subsubsection{Generator-based approach}
With the \emph{generator} network from our trained PGGAN as a ``black box'' simulator, random samples drawn from $p_G(\mathbf{x} | \mathbf{\theta})$ can be readily generated. Algorithm \ref{alg:ABC_a} lists schematically the steps needed to obtain an approximation of the posterior distribution $p(\mathbf{\theta} | \hat{\mathbf{x}})$. Due to the randomness of the sample generation, the discrepancy $\Delta_i$ is not a deterministic function of the parameter vector $\mathbf{\theta}_i$ (assuming fixed $\hat{\mathbf{x}}$), but rather a random variable depending on the noise vector $\mathbf{z}$. A new random noise sample should be drawn for each parameter vector $\theta_i$ in order to avoid a bias due to the distinct features emanating from a fixed noise sample. Possible choices for $\Delta$ are for instance a suitable norm between the power spectra or the PDFs. As $\varepsilon \searrow 0$, one expects the distribution defined by the accepted samples to gradually shift from the prior distribution $p(\mathbf{\theta})$ to the posterior distribution $p(\mathbf{\theta} | \hat{\mathbf{x}})$. Note that the approximation made for the posterior distribution is twofold, namely $p(\mathbf{\theta} | \hat{\mathbf{x}}) \approx p_G(\mathbf{\theta} | \hat{\mathbf{x}}) \approx p_G(\mathbf{\theta} | \Delta(\hat{\mathbf{x}}, \mathbf{x}) < \varepsilon)$, and the latter is approximated by a finite number of drawings of $\mathbf{\theta}$, which makes a careful sanity check of the obtained results mandatory. 

\begin{algorithm}
\caption{ABC Rejection Sampling: Generator-based}\label{alg:ABC_a}
\begin{algorithmic}[1]
\State Define $\varepsilon > 0$ sufficiently small
\For {($i = 0; \, i < N_{\text{samples}}; \, i$++)}
\State Draw $\mathbf{\theta}_i \sim p(\mathbf{\theta})$
\State Draw $\mathbf{z} \sim \mathcal{N}(\mathbf{0}, \boldsymbol {\mathsf{I}}_{N_\text{noise}})$
\State $\mathbf{x}_i \gets G(\mathbf{z}|\mathbf{\theta}_i) \sim p_G(\mathbf{x}|\mathbf{\theta}_i)$ 
\Comment{Generate PGGAN sample}
\State $\Delta_i \gets \Delta(\hat{\mathbf{x}}, \mathbf{x}_i)$
\If {$\Delta_i < \varepsilon$} 
\State Accept $\mathbf{\theta}_i$
\Else 
\State Reject $\mathbf{\theta}_i$
\EndIf
\EndFor
\State Approximate the posterior distribution $p(\mathbf{\theta} | \hat{\mathbf{x}})$ by the distribution of accepted parameters $\mathbf{\theta}_{i, \text{acc}}$
\end{algorithmic}
\end{algorithm}

\subsubsection{Critic-based approach}
For the \emph{generator}-based approach as described above, a suitable statistic needs to be defined based on which the proximity between a generated sample $\mathbf{x}_i$ and $\hat{\mathbf{x}}$ is measured. Furthermore, drawing the same $\mathbf{\theta}_i$ multiple times generally leads to different $\Delta_i$ owing to the randomness of the generated images. \orange{Another approach for parameter inference in the ABC framework relies on the trained \emph{critic} network of the PGGAN}: recalling that the \emph{critic} assesses the quality of each generated sample \emph{given an associated parameter vector}, one expects the \emph{critic} to assign a low score to a \emph{real} image paired with a wrong parameter vector. This motivates the use of the \emph{critic} score as a discrepancy measure for the ABC rejection sampling, as summarised in Algorithm \ref{alg:ABC_b}. Now, $\hat{\mathbf{x}}$ is shown to the \emph{critic} together with parameter vectors $\mathbf{\theta}_i$. Those that receive a high \emph{critic} score $\Delta_i > M$ are accepted, where $M \in \mathbb{R}$. Thus, there is no need to generate new samples and to manually define a statistic for the calculation of $\Delta_i$. Moreover, $\Delta_i$ depends deterministically on $\mathbf{\theta}_i$\footnote{The \emph{critic} score becomes stochastic if the \emph{critic} contains layers that introduce randomness at evaluation time, such as dropout layers.}. \orange{While the \emph{generator}-based approach is a classical application of an ABC technique for which a few theoretical results (e.g. \citealt{Dean2014}) and a solid amount of empirical studies exist, the \emph{critic}-based approach is clearly more heuristic and provides less interpretability since it is not transparent based on which criteria the \emph{critic} assigns a certain score to a tomographic sample, and in particular how the parameter vector enters this assessment. On the other hand, since the \emph{critic} has access to all the pixel values and not just a summary statistic, it is imaginable that the \emph{critic} is able to tell apart samples corresponding to different parameter vectors whose statistics such as power spectra are very similar.} 

\begin{algorithm}
\caption{ABC Rejection Sampling: Critic-based}\label{alg:ABC_b}
\begin{algorithmic}[1]
\State Define $M \in \mathbb{R}$
\For {($i = 0; \, i < N_{\text{samples}}; \, i$++)}
\State Draw $\mathbf{\theta}_i \sim p(\mathbf{\theta})$
\State $\Delta_i \gets D(\hat{\mathbf{x}} | \mathbf{\theta}_i)$ 
\Comment{Obtain \emph{critic} score}
\If {$\Delta_i > M$} 
\State Accept $\mathbf{\theta}_i$
\Else 
\State Reject $\mathbf{\theta}_i$
\EndIf
\EndFor
\State Approximate the posterior distribution $p(\mathbf{\theta} | \hat{\mathbf{x}})$ by the distribution of accepted parameters $\mathbf{\theta}_{i, \text{acc}}$
\end{algorithmic}
\end{algorithm}

\subsection{Inference example}
\begin{figure}
  \centering
  \noindent
  \resizebox{\columnwidth}{!}{
  \includegraphics{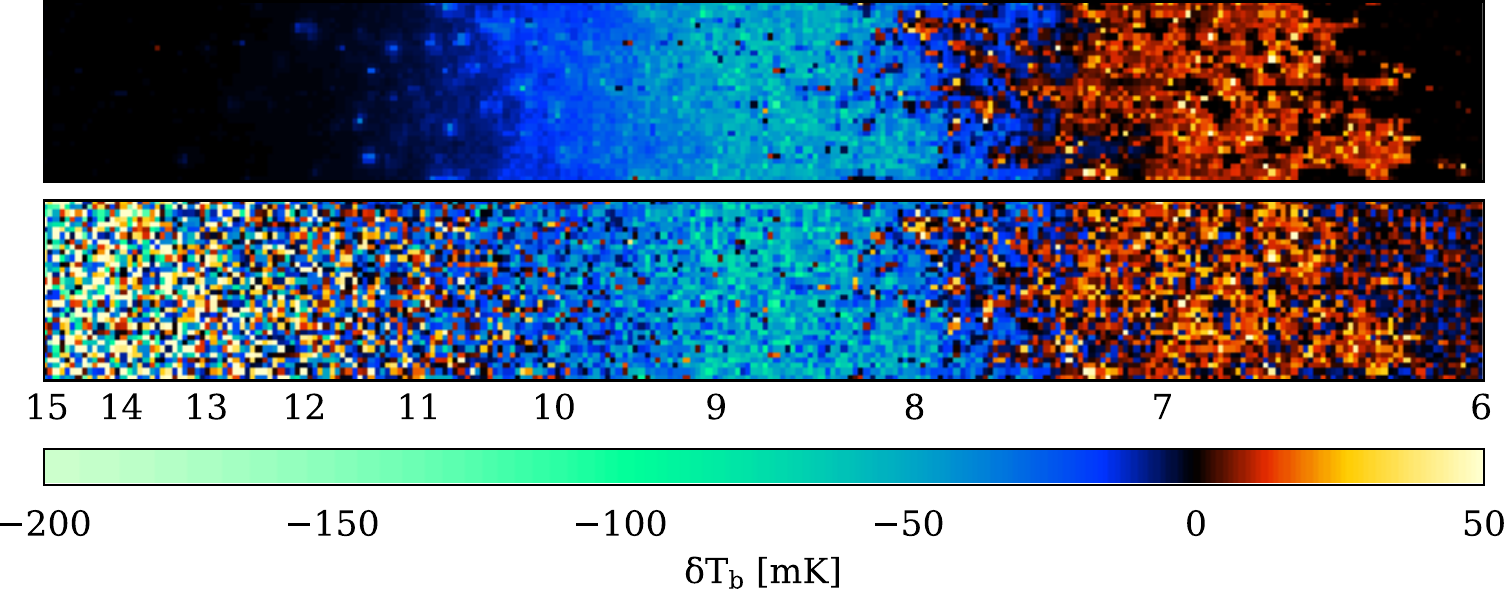}
  }
  \caption{\orange{The random 2D sample $\hat{\mathbf{x}}$ from the \citetalias{Semelin2017} catalogue for which we carry out the parameter inference, without noise (top) and with simulated thermal noise with SKA-like variance at each redshift (bottom). This sample belongs to the parameter vector $\hat{\mathcal{\theta}} = (\hat{f}_X, \hat{r}_{h/s}, \hat{f}_\alpha) = (1, 0.5, 1)$.  }}
  \label{fig:ABC_sample}
\end{figure}
We repeat the training of the PGGAN from Section \ref{sec:results}, but this time excluding all samples corresponding to the parameters $f_X = f_\alpha = 1$ from the training set (for all three values of $r_{h/s}$ since the effect of $r_{h/s}$ is less dominant than that of the other two parameters) in order to assess the ability of the PGGAN to interpolate to new points in parameter space, which is crucial for parameter inference.
\par As a test case, we take the sample $\hat{\mathbf{x}}$ shown in \orange{Figure \ref{fig:ABC_sample}}, which arises from a moderate reionisation history expressed \orange{by} the parameter vector $\hat{\mathbf{\theta}} = \orange{(\hat{f}_X, \hat{r}_{h/s}, \hat{f}_\alpha)} = (1, 0.5, 1)$. Since it is difficult to determine appropriate values for $\varepsilon$ and $M$ \emph{a priori}, we draw $49152$ samples \orange{$\mathbf{\theta}_i$} for each method and determine $\varepsilon$ and $M$ \emph{a posteriori} such that the number of accepted samples amounts to 512, which is roughly $1$ per cent. We checked that the resulting posterior distributions are sufficiently robust with respect to the accepted quantile, and accepting e.g. $0.5 - 2$ per cent instead of $1$ per cent gives similar results. 
\par In this example, we take non-informative priors over the parameter range spanned by the \citetalias{Semelin2017} catalogue. These are given by a uniform prior for $r_{h/s}$ and log-uniform priors for the scale variables $f_X$ and $f_\alpha$, i.e. $\log_{10}(f_X) \sim \mathcal{U}(-1, 1)$, $r_{h/s} \sim \mathcal{U}(0, 1)$, and $\log_{10}(f_\alpha) \sim \mathcal{U}(-\log_{10}(2), \log_{10}(2))$.
\par For the \emph{generator}-based approach, we take the discrepancy $\Delta$ to be the $1$-Wasserstein distance between the brightness temperature distributions at each scale factor $a = (1 + z)^{-1}$, integrated over the range of scale factors (or equivalently over the range of received frequencies). The scale factor $a$ is a proxy for the horizontal position within the tomography samples (recall that the cells in the samples have constant $\Delta \nu$ implying constant $\Delta a$). \orange{The motivation behind this choice of $\Delta$ is the following: for each scale factor, similar parameter vectors are expected to give similar PDFs. Binning the values of $\delta T_b$ at each scale factor, one could proceed by defining bin-to-bin discrepancy measures such as the absolute value of the difference between the PDFs in each temperature bin, summed over all bins. However, given that only $32$ pixel values of $\delta T_b$ are available at each scale factor for our 2D tomographic samples, very few values would be assigned to each temperature bin, and bin-to-bin measures would depend sensitively on the chosen temperature bin size. 
Contrarily, the Wasserstein distance as a cross-bin dissimilarity measure is much less sensitive to the chosen bin size: \purple{the amount of work required to move ``probability mass'' to an adjacent bin is small, and so is hence the resulting Wasserstein distance.}
To be more specific, the Wasserstein distance in the case of two Dirac delta distributions located at $T_1$ and $T_2$ collapses to $|T_1 - T_2|$ and simply measures the difference between the two temperatures. We take the $1$-Wasserstein distance and the $L^1$-norm over the scale factor for consistency with the loss function of the GAN. \purple{The reason for evaluating the Wasserstein distance between the PDFs at each scale factor instead of the PDFs of \emph{all} pixel values in the image is that two samples that have a very different evolution of the brightness temperature over time should be assigned a high discrepancy, even if the distribution of all pixel values taken in aggregate is similar.} In fact, binning $\delta T_b$ is not needed at all for the calculation of the $1$-Wasserstein distance, which can be done in 1D by means of a discrete form of the equivalent characterisation \citep{Vallender1974}:
\begin{equation}
    W_1(p_1, p_2) = \int_{-\infty}^\infty |P_1(x) - P_2(x)| \, \text{d}x, 
\end{equation}
where $P_1$ and $P_2$ are the cumulative distribution functions of $p_1$ and $p_2$, respectively.
Thus, we take}
\begin{equation}
\label{eq:Delta_i}
    \Delta_i = \Delta(\hat{\mathbf{x}}, \mathbf{x}_i) = \int_{a_\text{min}}^{a_\text{max}} W_1(\hat{t}(a), t_i(a)) \, \text{d}a,
\end{equation}
where $a_\text{min} = (1 + 15)^{-1}$, $a_\text{max} = (1 + 6)^{-1}$, \orange{and} $\hat{t} = \hat{t}(a)$ and $t_i = t_i(a)$ are the PDFs of $\hat{\mathbf{x}}$ and $\mathbf{x}_i$ at scale factor $a$, respectively\orange{.} 
The integrand is non-negative since the Wasserstein distance defines a metric. Numerically, we compute the $1$-Wasserstein distance between the brightness temperature distributions at each image column, and sum up the results over the columns. 

\subsection{Results}
\begin{figure}
  \centering
  \noindent
  \resizebox{\columnwidth}{!}{
  \includegraphics{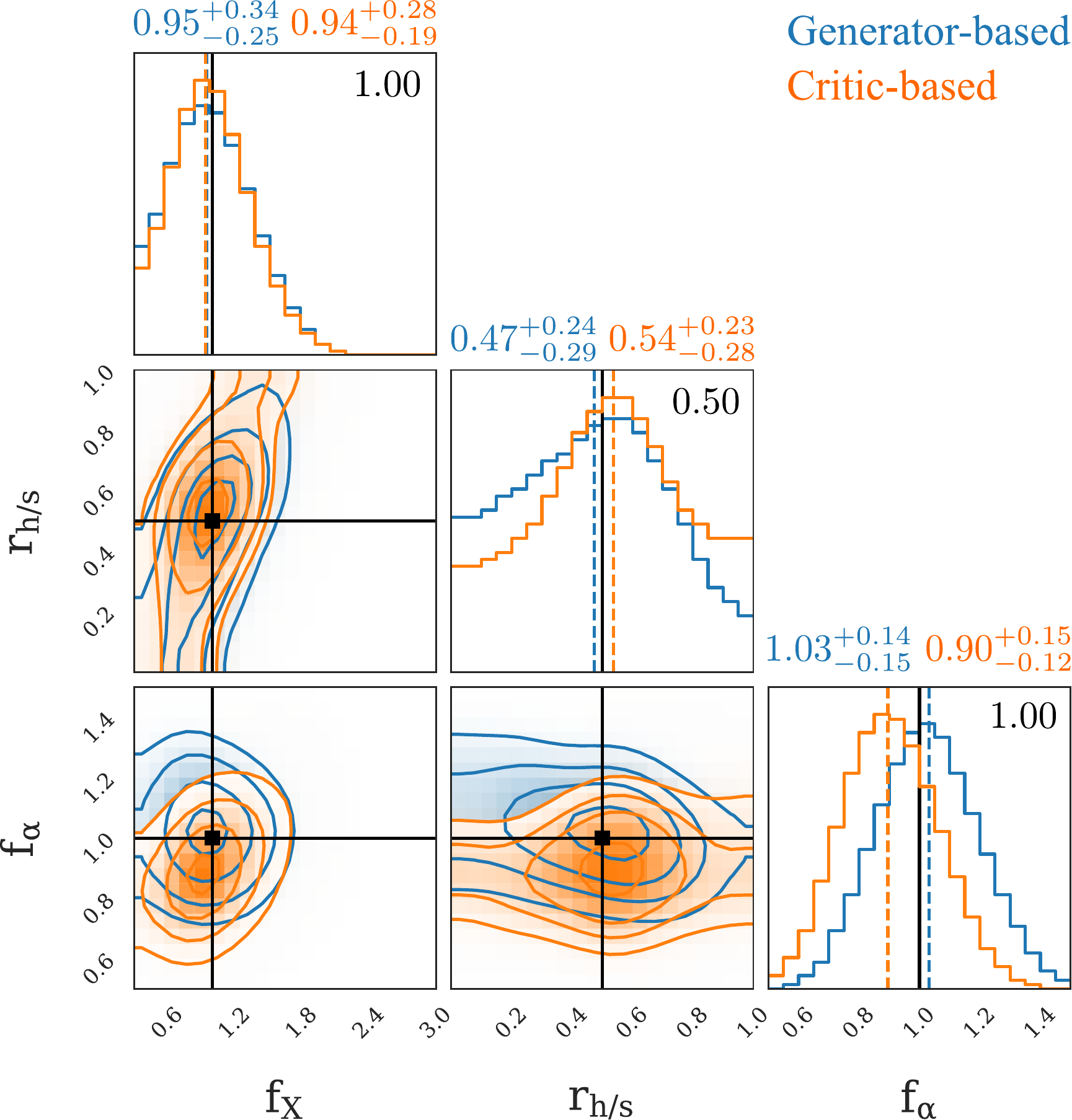}
  }
  \caption{Corner plot of the approximated posterior distribution using ABC rejection sampling with the \emph{generator}-based and \emph{critic}-based approach for the sample $\hat{\mathbf{x}}$ in \orange{Figure \ref{fig:ABC_sample} (without thermal noise)}. The true parameter values are given in black in the corners of the diagonal plots. We report the $68 \%$ equi-tailed credible intervals around the median in the plot titles. \orange{The contours in the 2D histograms mark the $0.5\sigma, 1\sigma, 1.5\sigma$, and $2\sigma$ regions.} A gentle Gaussian filter has been applied to the 1D and 2D histograms for presentation.}
  \label{fig:ABC_results}
\end{figure}
Figure \ref{fig:ABC_results} shows the resulting estimates of the posterior distribution $p(\mathbf{\theta}|\hat{\mathbf{x}})$ for the two different approaches. We stress again that for this PGGAN, all samples corresponding to $f_X = f_\alpha = 1$ were excluded from the training set, not only the particular sample $\hat{\mathbf{x}}$. This is because in a realistic scenario, it is unlikely that the underlying parameter vector of an observed tomography sample lies exactly on the parameter grid of the training set. 
\par Both approaches perform similarly well, although the \emph{generator}-based approach gives a somewhat more accurate estimate of the Lyman band emissivity $f_\alpha$. Given that the $(50 \pm 34) \%$ quantiles of the marginal prior distributions are $1.00^{+3.79}_{-0.79}$ and $1.00^{+0.60}_{-0.38}$ for $f_X$ and $f_\alpha$, respectively, the spread of the posterior distributions for $f_X$ and $f_\alpha$ is small (see the titles of the marginal distributions in Figure \ref{fig:ABC_results}), and the ABC inference has improved the bounds by a factor of a few. Determining the X-ray hard-to-soft ratio $r_{h/s}$ is harder due to its smaller imprint on the tomography samples as compared to the other two parameters (see Figure \ref{fig:interpolation}). The posterior distributions for both approaches peak close to the correct value $r_{h/s} = 0.5$, although the improvement with respect to the prior quantiles $0.50 \pm 0.34$ is small. This is in line with \citetalias{Semelin2017} who find that $r_{h/s}$ is difficult to constrain with the power spectrum and the PDF. The observed covariance between $r_{h/s}$ and $f_X$ is expected since increasing the X-ray emissivity and decreasing the X-ray hardness both results in a higher mean kinetic temperature of the gas.
\par For the \emph{generator}-based approach, our choice of $\Delta$ is only one out of many well-motivated discrepancy measures. We remark that replacing the integrand in Equation \eqref{eq:Delta_i} with the absolute difference between the mean brightness temperatures at scale factor $a$ barely changes the results. In order words, simply comparing the means for each scale factor instead of the full temperature distributions does not result in a 
loss in discriminatory power. However, statistics that capture more information such as the one proposed herein might be useful for future high-resolution 21cm imaging. 

\subsection{Inference in the presence of noise}
\label{subsec:noise}
Whereas we have assumed that the tomography sample with unknown parameters \orange{$\hat{\mathbf{x}}$} has the same quality as the training data in the inference above, real 21cm samples will be plagued by thermal noise and foregrounds that need to be removed. Therefore, it is vital to discuss how the above approaches carry over to a more realistic setting. For the \emph{generator}-based approach, accounting for noise is relatively straightforward: given an instrument-specific noise model, random noise realisations can be added to the GAN-generated samples before calculating $\Delta_i$. 
\orange{We repeat the generator-based parameter inference for the same sample, subject to Gaussian random noise with a redshift-dependent variance as it is expected from the SKA (depicted in Figure \ref{fig:ABC_sample}). We choose the same observational parameters as in \citetalias{Semelin2017}. For simplicity, we calculate the noise root mean square integrated over all wave numbers and subsequently draw random noise realisations in real space with SKA-like variance at each redshift (taking into account that the angular resolution for the fixed-size pixels varies with redshift), neglecting spatial correlations that would arise when carefully sampling the noise in the UV-plane and applying an inverse Fourier transform back to real space. 
\par Interestingly, the generator-based parameter inference barely deteriorates in the presence of noise: the resulting marginalised $68\%$ equi-tailed credibility regions are $0.93^{+0.36}_{-0.24}$ for $f_X$, $0.46^{+0.24}_{-0.29}$ for $r_{h/s}$, and $1.00^{+0.17}_{-0.14}$ for $f_\alpha$. This shows that astrophysical knowledge can be extracted in realistic situations where the measurements are impacted by both sample variance and thermal noise.} \purple{We emphasise again that even for upcoming 21cm surveys, thermal noise will be a greater source of uncertainty than sample variance, and one would expect the credibility regions to be set mainly by the noise. In such situations, the contribution of the epistemic uncertainty merits further investigation, which we will carry out in future work.}
\par On the other hand, dealing with noise in the \emph{critic}-based approach is more intricate: without introducing noise to the PGGAN, it is not clear whether the \emph{critic}, which has been trained on noiseless samples, will output high scores for a noisy image together with the correct parameter vector. A possible solution could be to add realistic noise to the generated samples during training before showing them to the \emph{critic}. While it is a common technique to add noise to the \emph{discriminator} input (or several layers of the \emph{discriminator}) in GANs \citep{Salimans2016} in order to make it harder to distinguish fake from real images, this often comes at the cost of reduced image quality. Another approach for both \emph{generator}-based and \emph{critic}-based inference could be adding noise to the training samples and \orange{training} the PGGAN to produce noisy samples. 
Our proposed methods exploit the fact that the \emph{generator} and \emph{critic} networks are trained anyway, but of course, it is also possible to employ an additional (possibly Bayesian) neural network for the task of parameter inference, as done in several references in the introduction of this work.

\section{Conclusions}
\label{sec:conclusions}
We have presented a framework for the fast creation of realistic 21cm tomography samples by means of a PGGAN. During training, the PGGAN learns to correctly account for X-ray emissivity $f_X$, X-ray hard-to-soft ratio $r_{h/s}$, and Lyman band emissivity $f_\alpha$, and to generate matching 21cm tomographic samples at gradually increasing resolution. Once trained, the PGGAN produces 2D samples at an SKA-like resolution of $\sim 3'$ ($32 \times 256$ pixels) in roughly a second on a laptop CPU. In comparison, the production of the \citetalias{Semelin2017} catalogue (which of course contains snapshots and lightcones at much higher resolution) took $3 \times 10^6$ CPU hours. The PGGAN-generated samples are diverse and accurately reproduce the global 21cm signal, the power spectrum, and the pixel distribution function of the training data. We have shown how both the \emph{generator} network and the \emph{critic} network can be harnessed for the task of parameter inference in the context of ABC rejection sampling. The reionisation parameters of an exemplary tomography sample are accurately recovered, and tight constraints are obtained for the parameters $f_X$ and $f_\alpha$ -- despite the fact that no samples for the correct parameter vector were contained in the training set. \orange{For the generator-based approach, this is even the case in the presence of simulated thermal noise with SKA-like variance, which shows the applicability of our approach in realistic scenarios.} 
In the dawning era of high-redshift 21cm imaging, deep learning techniques will provide valuable tools for forward and backward modelling. In this paper we have demonstrated how (PG)GANs can be utilised for both tasks. 

\section*{Acknowledgements}
The authors thank Beno\^{i}t Semelin for his help with the 21SSD \orange{data set and with the SKA noise computation}, and for making \orange{the data} publicly available. \orange{The authors also thank the anonymous referee for their feedback that improved the quality of this work.} The authors acknowledge the National Computational Infrastructure (NCI), which is supported by the Australian Government, for providing services and computational resources on the supercomputers Raijin \orange{and Gadi} that have contributed to the research results reported within this paper. This research made use of the Argus Virtual Research Desktop environment funded by the University of Sydney. F. L. is supported by the University of Sydney International Scholarship (USydIS).


\appendix
\section{Implementation details of the neural network}
\label{sec:NN_details}
Table \ref{table:PGGAN} lists the layers of the neural network at the final resolution. We found it beneficial in our experiments to replace pixel normalisation by instance normalisation \citep{Ulyanov2016}, except for directly after the input layer of the \emph{generator}. In contrast to \citetalias{Karras2017}, we process the \emph{generator} input with a fully connected layer instead of a convolutional layer. Besides, we noticed that the PGGAN becomes prone to mode collapse if the number of channels is reduced as the PGGAN grows deeper, which is why we keep the number of channels constant at $512$ which is feasible at the resolution considered herein -- however, for the creation of higher resolution samples, a bottleneck architecture might be attractive. Different from the common practice for GANs, downsampling in \citetalias{Karras2017} is not achieved by strided convolutions, but rather by average pooling, which we follow. The upsampling operation is a nearest neighbour interpolation. Recall that the channels of the \emph{critic} input correspond to $\delta T_b$, the global signal at each redshift $\langle\delta T_b\rangle$, $\delta T_b - \langle\delta T_b\rangle$, and the three model parameters. As suggested by \citetalias{Karras2017}, we employ an equalised learning rate and a minibatch standard deviation layer with group size 4. The decay rates of the moments for the Adam optimiser are taken to be $\beta_1 = 0.0$ and $\beta_2 = 0.99$.
\par It proved advantageous to take the logarithm of the two scale variables, i.e. X-ray and Lyman band emissivity $f_X$ and $f_\alpha$, respectively, and to normalise all components of $\mathbf{\theta}$ before feeding them to the neural network via the mapping $\theta_i \mapsto \frac{\theta_i - \bar{\theta}_i}{\sigma_i}$ for $i = 1, 2, 3$, where $\bar{\theta}_i$ and $\sigma_i$ are the mean and standard deviation of each parameter computed over the set of training images. 
\par For the training, we normalise the brightness temperatures via the mapping
\begin{equation}
    \delta T_b \mapsto \frac{\delta T_b}{175} + 0.4 + 0.075 \, \text{arcsinh}(0.5 \, \delta T_b).
\end{equation}
The inverse hyperbolic sine term leads to a steeper slope and hence to an increased sensitivity to small perturbations around $\delta T_b = 0$. The temperature range $[-177, 53] \ \text{mK}$ is monotonically mapped to $[-1, 1]$, and since we do not apply an activation function after the last $1 \times 1$ convolutional layer of the \emph{generator} that would confine the \emph{generator} output to a certain interval, the few pixels outside this range do not require any special treatment. We calculate the inverse transformation of the PGGAN output back to temperature space numerically.
\par The total number of trainable parameters for the fully grown PGGAN amounts to \orange{$\sim 2.65 \times 10^7$ and $\sim 2.60 \times 10^7$} for the \emph{generator} and the \emph{critic}, respectively. 

\begin{table}
\centering
\begin{tabular}{@{}ll@{}}
\toprule
\textbf{Operation}                         & \textbf{Output shape} ($C \times H \times W$)        \\ \midrule
\emph{GENERATOR}: & \\
\cmidrule(r){1-1}
Parameters and noise                       & 512 $\times$ 1 $\times$ 1    \\
IN $\circ$ LReLU $\circ$ FC $\circ$ PN     & 512 $\times$ 1 $\times$ 8    \\
IN $\circ$ LReLU $\circ$ Conv $1 \times 3$ & 512 $\times$ 1 $\times$ 8    \\ \cmidrule(r){1-1}
Upsampling                                 & 512 $\times$ 2 $\times$ 16   \\
IN $\circ$ LReLU $\circ$ Conv $3 \times 3$ & 512 $\times$ 2 $\times$ 16   \\
IN $\circ$ LReLU $\circ$ Conv $3 \times 3$ & 512 $\times$ 2 $\times$ 16   \\ \cmidrule(r){1-1}
Upsampling                                 & 512 $\times$ 4 $\times$ 32   \\
IN $\circ$ LReLU $\circ$ Conv $3 \times 3$ & 512 $\times$ 4 $\times$ 32   \\
IN $\circ$ LReLU $\circ$ Conv $3 \times 3$ & 512 $\times$ 4 $\times$ 32   \\ \cmidrule(r){1-1}
Upsampling                                 & 512 $\times$ 8 $\times$ 64   \\
IN $\circ$ LReLU $\circ$ Conv $3 \times 3$ & 512 $\times$ 8 $\times$ 64   \\
IN $\circ$ LReLU $\circ$ Conv $3 \times 3$ & 512 $\times$ 8 $\times$ 64   \\
\cmidrule(r){1-1}
Upsampling                                 & 512 $\times$ 16 $\times$ 128 \\
IN $\circ$ LReLU $\circ$ Conv $3 \times 3$ & 512 $\times$ 16 $\times$ 128 \\
IN $\circ$ LReLU $\circ$ Conv $3 \times 3$ & 512 $\times$ 16 $\times$ 128 \\
\cmidrule(r){1-1}
Upsampling                                 & 512 $\times$ 32 $\times$ 256 \\
IN $\circ$ LReLU $\circ$ Conv $3 \times 3$ & 512 $\times$ 32 $\times$ 256 \\
IN $\circ$ LReLU $\circ$ Conv $3 \times 3$ & 512 $\times$ 32 $\times$ 256 \\
Conv $1 \times 1$                          & 1   $\times$ 32 $\times$ 256 \\
\midrule
\emph{CRITIC}: & \\
\cmidrule(r){1-1}
Input tensor                               & 6   $\times$ 32 $\times$ 256 \\
LReLU $\circ$ Conv $1 \times 1$            & 512 $\times$ 32 $\times$ 256 \\
LReLU $\circ$ Conv $3 \times 3$            & 512 $\times$ 32 $\times$ 256 \\
LReLU $\circ$ Conv $3 \times 3$            & 512 $\times$ 32 $\times$ 256 \\
Downsampling                               & 512 $\times$ 16 $\times$ 128 \\
\cmidrule(r){1-1}
LReLU $\circ$ Conv $3 \times 3$            & 512 $\times$ 16 $\times$ 128 \\
LReLU $\circ$ Conv $3 \times 3$            & 512 $\times$ 16 $\times$ 128 \\
Downsampling                               & 512 $\times$ 8 $\times$ 64 \\
\cmidrule(r){1-1}
LReLU $\circ$ Conv $3 \times 3$            & 512 $\times$ 8 $\times$ 64 \\
LReLU $\circ$ Conv $3 \times 3$            & 512 $\times$ 8 $\times$ 64 \\
Downsampling                               & 512 $\times$ 4 $\times$ 32 \\
\cmidrule(r){1-1}
LReLU $\circ$ Conv $3 \times 3$            & 512 $\times$ 8 $\times$ 64 \\
LReLU $\circ$ Conv $3 \times 3$            & 512 $\times$ 8 $\times$ 64 \\
Downsampling                               & 512 $\times$ 4 $\times$ 32 \\
\cmidrule(r){1-1}
LReLU $\circ$ Conv $3 \times 3$            & 512 $\times$ 4 $\times$ 32 \\
LReLU $\circ$ Conv $3 \times 3$            & 512 $\times$ 4 $\times$ 32 \\
Downsampling                               & 512 $\times$ 2 $\times$ 16 \\
\cmidrule(r){1-1}
LReLU $\circ$ Conv $3 \times 3$            & 512 $\times$ 2 $\times$ 16 \\
LReLU $\circ$ Conv $3 \times 3$            & 512 $\times$ 2 $\times$ 16 \\
Downsampling                               & 512 $\times$ 1 $\times$ 8 \\
\cmidrule(r){1-1}
Minibatch STD                              & 513 $\times$ 1 $\times$ 8 \\
LReLU $\circ$ Conv $1 \times 4$            & 512 $\times$ 1 $\times$ 5 \\
LReLU $\circ$ Conv $1 \times 5$            & 512 $\times$ 1 $\times$ 1 \\
FC                                         & 1   $\times$ 1 $\times$ 1 \\
\bottomrule
\end{tabular}
\caption{Network architecture of the \emph{generator} and \emph{critic} for the fully grown PGGAN. PN: pixel normalisation, FC: fully connected layer, LReLU: leaky ReLU, IN: instance normalisation, Conv $k_h \times k_w$: convolutional layer with kernel size $k_h \times k_w$ and stride $1 \times 1$, Minibatch STD: minibatch standard deviation layer.}
\label{table:PGGAN}
\end{table}

\section{Samples for selected parameter vectors}
\label{sec:samples}
\begin{figure*}
  \centering
  \noindent
  \resizebox{0.73\textwidth}{!}{
  \includegraphics{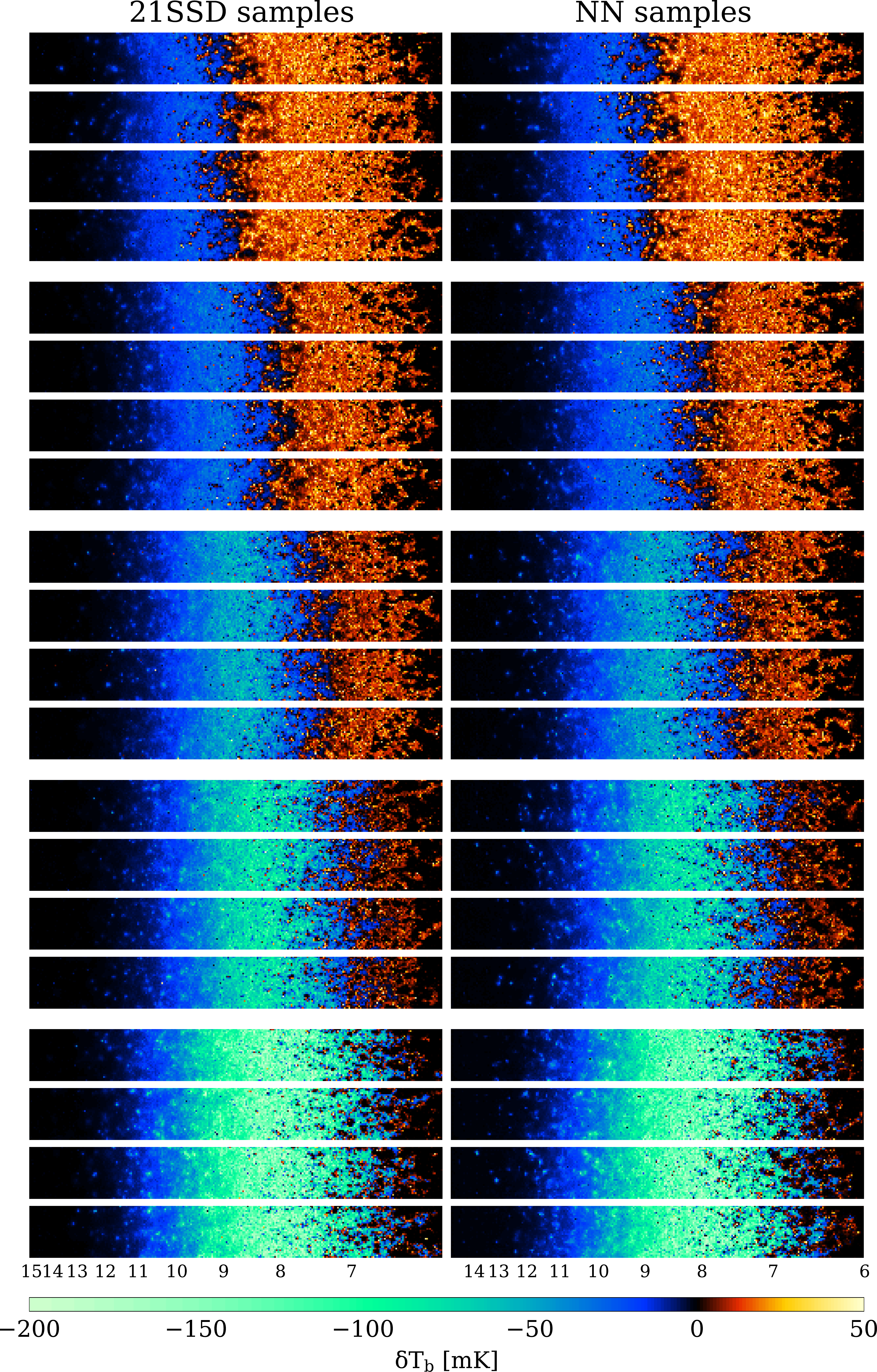}
  }
  \caption{Four random tomography samples from the \citetalias{Semelin2017} catalogue and from the PGGAN for each of the following five parameter sets: $(f_X, r_{h/s}, f_\alpha) = (10, 0, 2)$, $(3, 0.5, 1)$, $(1, 0.5, 1)$, $(0.3, 0.5, 1)$, $(0.1, 1, 2)$, from top to bottom.}
  \label{fig:fixed_samples}
\end{figure*}
Figure \ref{fig:fixed_samples} shows four random \citetalias{Semelin2017} and PGGAN samples for each of five different parameter choices. The quality of the PGGAN samples is high, and each sample features an individual distribution of the fluctuations.

\bsp	
\label{lastpage}
\end{document}